\documentclass[aps, prl, floatfix,
  reprint, 
  superscriptaddress, 
  longbibliography, noeprint,
]{revtex4-2} 
\usepackage[T1]{fontenc}
\usepackage[utf8]{inputenc}
\usepackage{microtype}
\usepackage{graphicx, xcolor} 
\usepackage[notrig]{physics} 
\usepackage{amsmath, amssymb}
\usepackage{mathtools} 
\usepackage{csquotes} 
\usepackage{siunitx} 
\usepackage[pdftex]{hyperref}
\hypersetup{
    colorlinks=true,
    citecolor=blue,
    linkcolor=magenta,
    urlcolor=blue,
    bookmarksnumbered=true,
    pdfborder={0 0 0},
    bookmarkstype=toc,
    pdfauthor={Shoki Sugimoto, Ryusuke Hamazaki, and Masahito Ueda}
}

\usepackage{amsthm} 
\newtheorem{conjecture}{Conjecture}
\newtheorem{definition}{Definition}
\newtheorem{proposition}{Proposition}
\renewcommand{\qed}{$\hfill\blacksquare$}

\usepackage{xparse}
\NewDocumentCommand{\seminorm}{s O{\opSet} m m}{%
  \IfBooleanTF{#1}
    {\norm*{#4}_{#3}^{(#2)}}%
    {\norm{#4}_{#3}^{(#2)}}%
}
\newcommand{\ETHmeasure}[3][1]{ \Lambda_{#1}^{(#2, #3)} }

\newcommand{\dimLoc}{ d_{\mathrm{loc}} }
\newcommand{\particleNumber}{ N }
\newcommand{\sysSize}{ V }
\newcommand{\EigStateOp}[1]{ \hat{\rho}_{#1} }
\newcommand{\MCE}[1][]{ \hat{\rho}^{(\mathrm{mc})}_{#1} } 
\newcommand{\opRange}[2][1]{ \norm*{ #2 }_{#1} }
\newcommand{\opSet}{ \mathcal{A} }
\newcommand{\Prob}{\mathbb{P}}
\newcommand{\upperBound}[1][]{#1{\alpha}_{\mathrm{U}}}
\newcommand{\lowerBound}[1][]{#1{\alpha}_{\mathrm{L}}}
\newcommand{\Supple}{Supplementary Material}

\DeclareMathOperator{\vecspan}{span}

\usepackage[resetlabels]{multibib}
\newcites{SM}{References}
\let\origciteSM\citeSM
\RenewDocumentCommand{\citeSM}{O{} O{} m}{%
  \begingroup
  \renewcommand{\citenumfont}[1]{S##1}%
  \origciteSM[#1][#2]{#3}%
  \endgroup
}
\let\origbibliographySM\bibliographySM
\renewcommand{\bibliographySM}[1]{%
  \begingroup
    \def\bibnumfmt##1{[S##1]}%
    \def\@biblabel##1{[S##1]}%
    \origbibliographySM{#1}%
  \endgroup
}
\usepackage{etoolbox}
\makeatletter
  \robustify\citeSM
\makeatother

\makeatletter
\def\frontmatter@maketitle{%
  \@author@finish
  \title@column\titleblock@produce
  \suppressfloats[t]%
  \let\abstract\@undefined\let\endabstract\@undefined
  \titlepage@sw{%
   \vfil
   \clearpage
  }{}%
  \onecolumn@grid@setup
  \def\set@footnotewidth{\set@footnotewidth@one}%
}%
\makeatother
\usepackage{tcolorbox} 
\tcbuselibrary{theorems, skins, breakable, hooks}
\tcbset{thm/.style={breakable, enhanced, drop fuzzy shadow, boxrule=1pt, colback=lightgray!15,colframe=gray!90,colbacktitle=blue!10,coltitle=black,fonttitle=\bfseries\strut}}
\newtcbtheorem[number within=section]{framedDefinition}{Definition}{thm}{defSM}
\newtcbtheorem[number within=section]{framedProposition}{Proposition}{thm}{propSM}
\newtcbtheorem[number within=section]{framedTheorem}{Theorem}{thm}{thSM}
\newtcbtheorem[number within=section]{framedLemma}{Lemma}{thm}{lemSM}
\newcommand{\subSystem}{ \mathcal{S} }

\begin{document}
\title{
    Bounds on eigenstate thermalization
}
\author{Shoki Sugimoto}
  \email{shoki.sugimoto@ap.t.u-tokyo.ac.jp}
  \affiliation{Department of Applied Physics, The University of Tokyo, 7-3-1 Hongo, Bunkyo-ku, Tokyo 113-0033, Japan}
  \affiliation{Nonequilibrium Quantum Statistical Mechanics RIKEN Hakubi Research Team, RIKEN Pioneering Research Institute (PRI), Wako, Saitama 351-0198, Japan}
\author{Ryusuke Hamazaki}
  \affiliation{Nonequilibrium Quantum Statistical Mechanics RIKEN Hakubi Research Team, RIKEN Pioneering Research Institute (PRI), Wako, Saitama 351-0198, Japan}
  \affiliation{RIKEN Center for Interdisciplinary Theoretical and
Mathematical Sciences (iTHEMS), RIKEN, Wako 351-0198, Japan
}
\author{Masahito Ueda}
  \affiliation{Department of Physics, The University of Tokyo, 7-3-1 Hongo, Bunkyo-ku, Tokyo 113-0033, Japan}
  \affiliation{RIKEN Fundamental Science Program (FQSP), Wako 351-0198, Japan}
\begin{abstract}
    The eigenstate thermalization hypothesis~(ETH), which asserts that every eigenstate of a many-body quantum system is indistinguishable from a thermal ensemble, plays a pivotal role in understanding thermalization of isolated quantum systems.
    Yet, no evidence has been obtained as to whether the ETH holds for \textit{all} few-body operators in a chaotic system; such few-body operators include key quantities in statistical mechanics, such as the total magnetization, the momentum distributions, and their low-order thermal and quantum fluctuations.
    %
    Here, we formulate a conjecture that for a generic nonintegrable system the ETH holds simultaneously for all $m$-body operators with $m < \alpha_{\ast} \particleNumber$ in the thermodynamic limit for some nonzero constant $\alpha_{\ast} > 0$.
    We first show the existence of such nontrivial constants for idealized (pseudo) random-matrix descriptions of many-body eigenstates.
    We then verify the conjecture for generic spin, Bose, and Fermi systems with local and few-body interactions by large-scale numerical calculations.
    Our results imply that generic systems satisfy the ETH simultaneously for \textit{all} few-body operators, including their thermal and quantum fluctuations.
\end{abstract}
\maketitle
\addtocontents{toc}{\string\tocdepth@munge} 

%
Recent experiments in cold atoms and ions have demonstrated that quantum systems thermalize unitarily without heat reservoirs~\cite{trotzky2012probing, Langen2013-zj, clos2016time, kaufman2016quantum, neill2016ergodic, tang2018thermalization, orioli2018relaxation, langen2015ultracold}.
This discovery brings up a striking possibility of integrating statistical mechanics into a single framework of quantum mechanics---a scenario first envisioned by von Neumann about a century ago~\cite{von2010proof}.
It has been argued that a single pure quantum state is indistinguishable from a thermal ensemble for few-body observables~\cite{popescu2006entanglement, goldstein2006canonical, reimann2007typicality,
tasaki2016typicality, polkovnikov2011colloquium, d2016quantum, gogolin2016equilibration, mori2018thermalization, deutsch2018eigenstate} due to the interplay between quantum entanglement and physical constraints on observables such as locality or few-bodiness~\cite{popescu2006entanglement, goldstein2006canonical, polkovnikov2011colloquium, d2016quantum, gogolin2016equilibration, mori2018thermalization, deutsch2018eigenstate}.
In fact, different notions of thermal equilibrium, such as microscopic thermal equilibrium~(MITE) and macroscopic thermal equilibrium~(MATE), can be introduced~\cite{popescu2006entanglement, goldstein2006canonical, Goldstein2015, Goldstein2017, tasaki2016typicality, mori2018thermalization}, depending on the nature of operations used to distinguish a quantum state from a thermal ensemble.

The eigenstate thermalization hypothesis (ETH)~\cite{von2010proof, deutsch1991quantum, srednicki1994chaos} is widely believed to be the primary mechanism behind thermalization in isolated quantum systems.
The ETH for an operator $\hat{A}$ means that (i) every energy eigenstate of a system is in thermal equilibrium with respect to the expectation value of $\hat{A}$ and that (ii) off-diagonal elements of $\hat{A}$ in an energy eigenbasis are vanishingly small.
The ETH ensures thermalization of $\hat{A}$ for any initial state with a macroscopically definite energy, barring massive degeneracy in the energy spectrum~\cite{polkovnikov2011colloquium, d2016quantum, gogolin2016equilibration, mori2018thermalization, deutsch2018eigenstate}.
The ETH has been tested numerically for several local or few-body quantities~\cite{rigol2008thermalization, beugeling2014finite, kim2014testing, sugimoto2021test, Sugimoto2022, rigol2009quantum, rigol2009breakdown, rigol2010quantum, biroli2010effect, santos2010localization, steinigeweg2013eigenstate, mondaini2016eigenstate, mondaini2017eigenstate, jansen2019eigenstate, Schonle_PhysRevB2021y}.
However, whether the ETH holds for \textit{all} few-body operators and whether it breaks down for many-body operators have yet to be fully understood.

Von Neumann~\cite{von2010proof} and Reimann~\cite{reimann2015generalization} proved the ETH for \textit{almost all} Hermitian operators.
However, it has yet to be established whether the ETH holds simultaneously for \textit{all} elements of a set of physically relevant operators, such as few-body observables.
Some works~\cite{garrison2018does, mori2018thermalization, dymarsky2018subsystem} tested the ETH against all operators supported on local subsystems; however, these approaches are inherently limited to subsystem observables and hence not designed to handle generic few-body operators that act on the entire system, such as the total magnetization, the momentum distribution, and their thermal and quantum fluctuations.

\begin{figure*}[t]
    \centering
    \includegraphics[width=\linewidth]{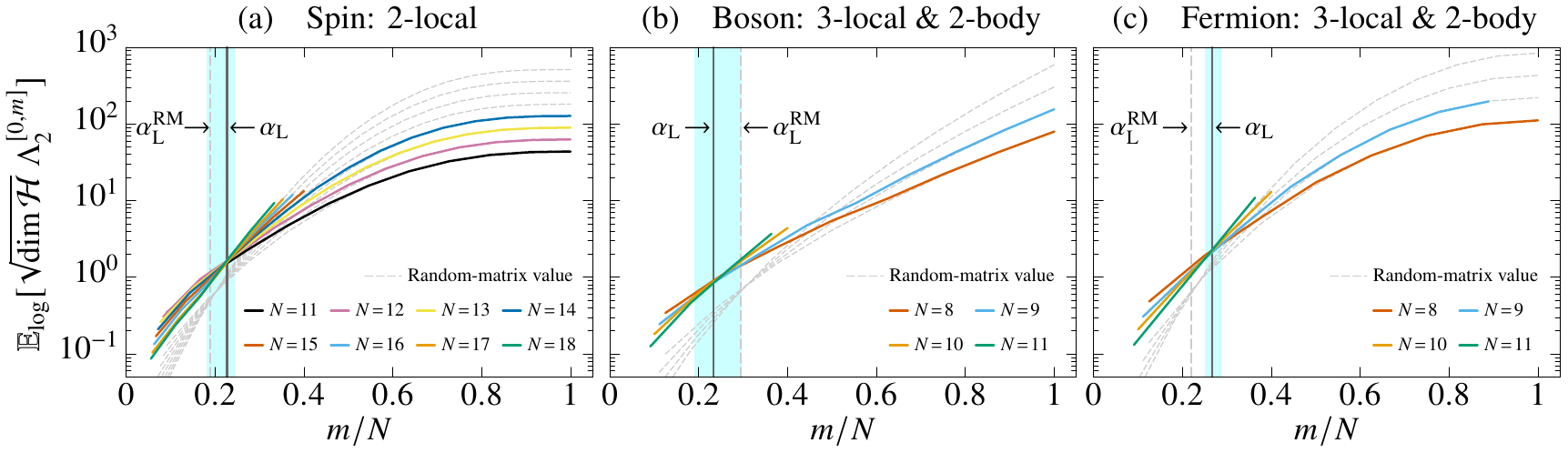}
    \caption{\textbf{Ensemble average of the upper bound on the diagonal ETH measure for local and few-body Hamiltonians.}
    $\Lambda_{2}^{[0,m]}$ is evaluated in the central energy window $E=(E_{\mathrm{max}}+E_{\mathrm{min}})/2$ with width $2\Delta\!{E}=0.05(E_{\mathrm{max}}-E_{\mathrm{min}})$ and microcanonical width $\var{E}=0.2(E_{\mathrm{max}}-E_{\mathrm{min}})/\sysSize$; $\mathbb{E}_{\mathrm{log}}$ denotes the geometric mean.
    The black vertical line gives $\lowerBound{}$, defined by the minimum of the maximum finite-size difference, and the blue shaded region estimates its finite-size uncertainty from the change between decreasing and increasing $\particleNumber$-dependence.
    The gray dashed curves show the upper bound in Eq.~\eqref{eq_RandomMatrixEstimate} for an idealized random-matrix model, whose intersection gives $\lowerBound^{\mathrm{RM}}$.
    Panels show (a) spin systems with up to 2-local terms and (b,c) Bose/Fermi systems with 3-local two-body terms.
    Colors denote system sizes.
    The finite-size crossings and the decrease of the upper bound to their left support Conjecture~\ref{conj_Conjecture} for Eqs.~\eqref{eq_SpinHamiltonian} and \eqref{eq_BosonFermionHamiltonian}, giving a nontrivial lower bound $\lowerBound{}>0$ on $\alpha_{\ast}$.}
    \label{fig_Numerics_ShortRangeEnsemble}
\end{figure*}

\begin{figure*}[t]
    \centering
    \includegraphics[width=\linewidth]{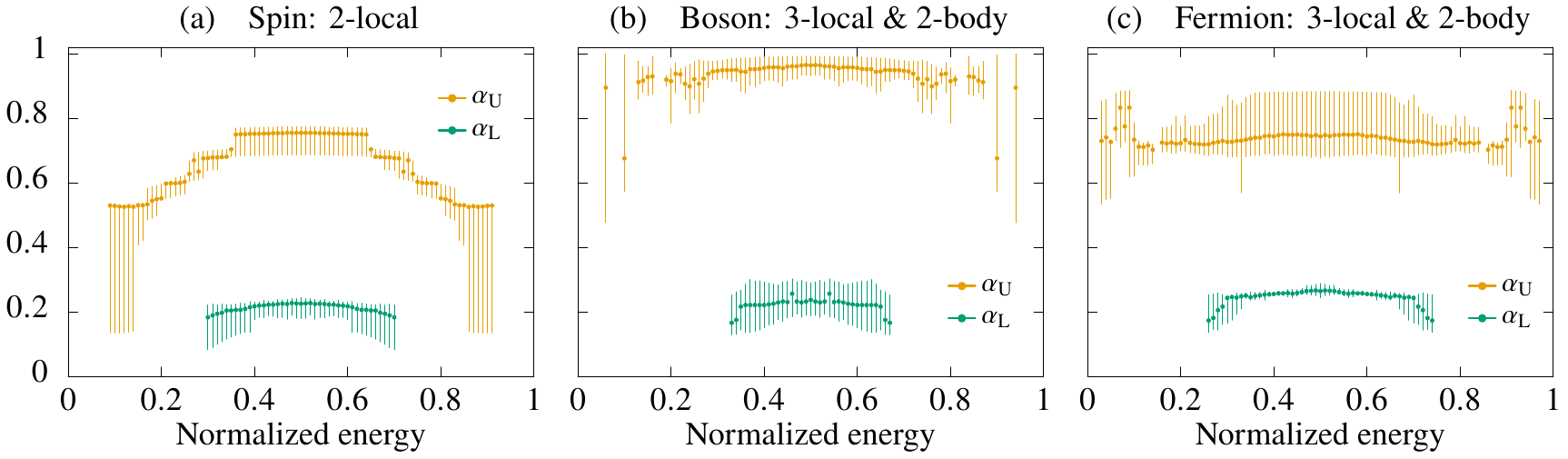}
    \caption{\textbf{Energy dependence of the bounds $\lowerBound$ and $\upperBound$ on the threshold $\alpha_{\ast}$.
    }
    The data points indicate the value of $m/\particleNumber$ where the maximum difference between the curves with different system sizes for a fixed normalized energy $x \coloneqq (E-E_{\mathrm{min}})/(E_{\mathrm{max}} - E_{\mathrm{min}})$, as shown in Fig.~\ref{fig_Numerics_ShortRangeEnsemble}, becomes minimal.
    The upper/lower end of the error bars shows the value of $\alpha$ above/below which $\mathbb{E}_{\mathrm{log}}[\sqrt{\dim\mathcal{H}}\, \Lambda_{2}^{[0,\alpha \particleNumber]}]$ (or $\mathbb{E}_{\mathrm{log}}[\Lambda_{2}^{[0,\alpha \particleNumber]}]$) monotonically increases/decreases as the system size increases.
    The system sizes used in this figure are $12\:(12) \leq \particleNumber \leq 18\:(14)$ for $\lowerBound{}\:(\upperBound{})$ of spin systems and $8\:(6) \leq \particleNumber \leq 11\:(9)$ for Bose and Fermi systems.
    The existence of the nontrivial lower bound $\lowerBound{}\, (>0)$ across approximately the central 40\% of the spectrum supports the validity of Conjecture~\ref{conj_Conjecture} in this energy region.
    }
    \label{fig_BoundsEnergyDependence}
\end{figure*}

In this Letter, we submit a conjecture that the ETH holds true for \textit{all} operators up to $\order{N}$-body ones.
More precisely, we formulate the following conjecture:
\begin{conjecture} \label{conj_Conjecture}
    For a generic nonintegrable system, 
    there exists a positive constant $\alpha > 0$ such that the ETH holds simultaneously for all $m$-body operators with $m < \alpha \particleNumber$ in the thermodynamic limit.
\end{conjecture}
\noindent
Let $\alpha_\ast$ be the largest such constant.
We first examine this conjecture for an idealized (pseudo)random-matrix description of chaotic many-body eigenstates~\cite{von2010proof, reimann2015generalization, Pappalardi_PhysRevLett2022z}.
We then test Conjecture~\ref{conj_Conjecture} numerically for generic spin, Bose, and Fermi systems with local and few-body interactions, employing finite-size scaling of the bounds on the ETH measure~(Figure~\ref{fig_Numerics_ShortRangeEnsemble}).
We compare the behavior of these numerical bounds, as we increase $N$ with $m/N$ held fixed, to the corresponding analytic behavior derived for idealized random-matrix descriptions of many-body eigenstates.
The obtained agreement supports extrapolating the numerical results to the thermodynamic limit.
This yields nontrivial lower bounds on $\alpha_{\ast}$ across approximately the central 40\% of the energy spectrum (Figure~\ref{fig_BoundsEnergyDependence}), thereby verifying Conjecture~\ref{conj_Conjecture} for the systems under investigation.

These results suggest that our method remains effective even in analyzing systems with non-negligible finite-size effects, such as trapped ions and cold atoms, where the thermodynamic limit is inaccessible~\cite{trotzky2012probing, Langen2013-zj, clos2016time, kaufman2016quantum, neill2016ergodic, tang2018thermalization, gring2012relaxation, neyenhuis2017observation}.
Moreover, our result applies directly to any few-body operator of interest in statistical mechanics, including their thermal and quantum fluctuations, without relying on coarse-graining or other approximations.
In particular, the ETH is expected to hold even for \textit{arbitrary} low-order fluctuations of \textit{arbitrary} few-body operators.

\paragraph{Unified measure for quantum-thermal equilibrium.---}
While several measures and criteria of quantum-thermal equilibrium have been introduced in the literature~\cite{Goldstein2015, Goldstein2017, tasaki2016typicality, mori2018thermalization}, 
they are specifically formulated for particular classes of operators (such as subsystem observables and macroscopic observables) and not applicable to generic few-body operators.

To overcome this limitation and to quantitatively study how the ETH depends on physical constraints on observables, 
we introduce the following measure of the closeness to a thermal ensemble, which is applicable to an arbitrary set $\opSet$ of observables.
\begin{definition}[Unified measure of quantum-thermal equilibrium]
    To quantify the distance $\norm*{ \hat{\sigma} - \hat{\rho}_{\mathrm{th}} }$ between a quantum state $\hat{\sigma}$ and a thermal state $\hat{\rho}_{\mathrm{th}}$, we introduce the following (semi-)norm,
    \begin{equation}
        \seminorm*{p}{ \hat{X} }
        \coloneqq \sup_{ \substack{ \hat{A} \in \opSet + \mathbb{R} \hat{I} \\ \hat{A} \neq 0}} 
        \abs{ \tr\qty( \frac{\hat{A}^{\dagger}}{ \opRange[q]{\hat{A}} } \,\hat{X} )  }, 
        \label{def_pseudoDistance1}
    \end{equation}
    where $p^{-1} +q^{-1} = 1$, $\mathbb{R}$ denotes the field of real numbers, and $\hat{I}$ is the identity operator.
    Here, $\norm*{ \hat{A} }_{q} \coloneqq (\tr[(\hat{A}^{\dagger}\hat{A})^{q/2})])^{1/q}$ is the Schatten-$q$ norm~\cite{bhatia2013matrix}.
    The operator $\hat{X}$ is arbitrary and not necessarily Hermitian.
    Then, the (pseudo-)distance $\seminorm{1}{ \hat{\sigma} - \hat{\rho}_{\mathrm{th}} }$ with $\hat{\rho}_{\mathrm{th}}$ being a thermal ensemble serves as a unified measure of quantum-thermal equilibrium.
\end{definition}
In the following, we only consider the cases $(p,q) = (1,\infty)$ and $(2,2)$ due to the following two reasons:
First, among $\seminorm{p}{\cdot}$ with $p\geq 1$, only $\seminorm{1}{\cdot}$ can be used to define a measure of quantum-thermal equilibrium, as it satisfies the following essential properties~\cite{sugimoto2021test}:
(i) invariant under a linear transformation$\colon \hat{A} \mapsto a' \hat{A} +b'$,
(ii) dimensionless, and
(iii) thermodynamically intensive, meaning that $\expval*{\hat{A}}^{(\mathrm{mc})}(E) / \norm*{\hat{A}}_{q}$ converges to a finite value in the thermodynamic limit.
Here, $\expval*{\hat{A}}^{(\mathrm{mc})}(E)$ denotes the microcanonical average at energy $E$.
Indeed, for an additive operator $\hat{A}$, we have $\norm*{\hat{A}}_{\infty} \sim \particleNumber$~\cite{Tasaki2018-bz}, ensuring the property (iii) for the 1-norm $\seminorm{1}{\cdot}$.
In contrast, we have $\norm*{\hat{A}}_{q} \sim (\dim \mathcal{H})^{1/q} \sqrt{N}$ for any $q \in [1,\infty)$~(see the End Matter).
Second, for general $\opSet$, it is difficult or even impossible to calculate the supremum in $\seminorm{1}{\cdot}$, both numerically and analytically.
However, the 2-norm $\seminorm{2}{\cdot}$ is computable for a given orthonormal basis of $\opSet + \mathbb{R}\hat{I}$ and bounds the 1-norm $\seminorm{1}{\cdot}$ from below and above~(see the End Matter for the proof).
\begin{proposition} \label{Proposition1}
    For any operator space $\opSet$ and any operator $\hat{X}$ not necessarily Hermitian, the 1-norm is bounded by the 2-norm from below and above as
    \begin{equation}
        \seminorm*{2}{\hat{X}} \leq \seminorm*{1}{\hat{X}} \leq \sqrt{\dim \mathcal{H}}\ \seminorm*{2}{\hat{X}}.
        \label{eq_BoundsOn1norm}
    \end{equation}
    Moreover, given an orthonormal basis $\Bqty*{ \hat{\Lambda}_{\ell} }_{\ell}$ of $\opSet + \mathbb{R} \hat{I}$, the 2-norm can be computed as
    \begin{equation}
        \seminorm*{2}{\hat{X}} = \sqrt{ \frac{ \norm*{ \vec{X} }_{2}^{2} + \abs*{ \vec{X}^{T} \cdot \vec{X} } }{2} },\quad (\vec{X})_{\ell} \coloneqq \tr( \hat{\Lambda}_{\ell}^{\dag} \hat{X} ).
        \label{eq_ComputableFormulaFor2Norm}
    \end{equation}
\end{proposition}

We say that a quantum state $\hat{\sigma}$ is in thermal equilibrium relative to $\opSet$ if $\seminorm{1}{ \hat{\sigma} - \hat{\rho}_{\mathrm{th}} } < \epsilon$ holds for a sufficiently small $\epsilon\, (>0)$.
This definition follows Refs.~\cite{Goldstein2015, Goldstein2017}; however, we here concern only the expectation value of $\hat{A} \in \opSet$ and not the probability distribution over the spectrum of $\hat{A}$.
Nonetheless, our framework can deal with the probability distribution by including sufficiently high powers of $\hat{A}$ in $\opSet$, offering finer control over the precision in observing the distribution of $\hat{A}$.
By appropriately choosing $\opSet$, our notion of quantum-thermal equilibrium unifies previously introduced notions of thermal equilibrium, such as subsystem thermal equilibrium~\cite{popescu2006entanglement, goldstein2006canonical, garrison2018does, dymarsky2018subsystem}, microscopic thermal equilibrium~(MITE), and macroscopic thermal equilibrium~(MATE)~\cite{Goldstein2015, Goldstein2017, mori2018thermalization}~(see \Supple{}~\ref{SectionSM1}).

Many of the previous works that numerically tested the ETH with respect to several operators $\hat{A}_{1},\cdots, \hat{A}_{J}$~\cite{rigol2008thermalization, beugeling2014finite, kim2014testing, rigol2009quantum, rigol2009breakdown, rigol2010quantum, biroli2010effect, santos2010localization, steinigeweg2013eigenstate, mondaini2016eigenstate, mondaini2017eigenstate, jansen2019eigenstate, Schonle_PhysRevB2021y} essentially set $\opSet = \Bqty*{ \hat{A}_{1},\cdots, \hat{A}_{J} }$.
However, if one really wants to distinguish an energy eigenstate from $\hat{\rho}_{\mathrm{th}}$ without any exception, one should use not only a limited number of quantities but \textit{all} the quantities compatible with physical constraints under consideration.

\paragraph{Measure of the ETH.---}
As the measure of the ETH, which requires (i) \textit{all} energy eigenstates to be in thermal equilibrium and (ii) \textit{all} off-diagonal elements of an observable in an energy eigenbasis to be vanishingly small, we introduce
\begin{equation*}
    \ETHmeasure[p]{\hat{H}}{\opSet}(E)
    \coloneqq \max_{\ket*{E_{\alpha}}, \ket*{E_{\beta}} \in \mathcal{H}_{E,\Delta\!{E}} }
    \seminorm{p}{ \EigStateOp{\alpha\beta} - \MCE[\var{E}](E_{\alpha}) \delta_{\alpha\beta} },
\end{equation*}
where $p=1$ or $2$, $\EigStateOp{\alpha\beta} \coloneqq \dyad*{ E_{\alpha} }{ E_{\beta} }$ with $\ket*{ E_{\alpha} }$ being an energy eigenstate with eigenenergy $E_{\alpha}$, and $\MCE[\var{E}](E_{\alpha})$ is the microcanonical density operator within the energy shell $\mathcal{H}_{E_{\alpha}, \var{E}} \coloneqq \vecspan\Bqty*{ \ket*{E_{\gamma}} \mid \abs*{E_{\gamma} - E_{\alpha}} < \var{E} }$.
Here, the shell width $\var{E}$ should be chosen sufficiently small as $\var{E} = \order*{N g(N)}$ when testing the ETH at resolution $\ETHmeasure[1]{\hat{H}}{\opSet}(E) = \order{g(N)}$, where $g(N)$ is a function that vanishes for $N\rightarrow\infty$.
As with the norm $\seminorm{p}{\cdot}$, $\ETHmeasure{\hat{H}}{\opSet}$ is a proper measure of the ETH, and $\ETHmeasure[2]{\hat{H}}{\opSet}$ will be used to bound $\ETHmeasure{\hat{H}}{\opSet}$:
\begin{equation}
    \ETHmeasure[2]{\hat{H}}{\opSet} \leq \ETHmeasure{\hat{H}}{\opSet} \leq \sqrt{\dim\mathcal{H}}\, \ETHmeasure[2]{\hat{H}}{\opSet}
    \label{eq_BoundsOnETHmeasure}
\end{equation}
We say that the ETH holds for an observable $\hat{A}$ if and only if $\lim_{N \to \infty} \ETHmeasure{\hat{H}}{\Bqty*{\hat{A}}} = 0$. 
Because the leading order in the thermodynamic limit is given by $\norm*{\hat{A}}_{\infty}$, this condition is equivalent to requiring that (i) the expectation values of $\hat{A}$ in $\ket{E_\alpha}$ and $\MCE[\var{E}](E_\alpha)$ agree at the leading order, and (ii) the off-diagonal elements of $\hat{A}$ are negligible at the same order.
This formulation aligns with the standard interpretation in statistical mechanics, whereby thermal ensembles that agree at leading order are regarded as equivalent, even if they differ at subleading orders~\cite{Ruelle1999-sb, Tasaki2018-bz}.
Accordingly, $\ETHmeasure{\hat{H}}{\opSet}$ detects both the validity and breakdown of the ETH in a manner consistent with previous numerical results~(see \Supple{}~\ref{SectionSM2}).
Since $\ETHmeasure{\hat{H}}{\opSet} = \sup_{\hat{A} \in \opSet+\mathbb{R}\hat{I}} \ETHmeasure{\hat{H}}{\Bqty*{\hat{A}}}$, the ETH holds for \textit{all} operators in $\opSet$ if and only if $\lim_{\particleNumber \to \infty} \ETHmeasure{\hat{H}}{\opSet} = 0$.
In particular, any previously reported breakdown of the ETH for specific observables~(e.g., \cite{rigol2009breakdown, santos2010localization, Turner_PhysRevB2018k}) also implies its breakdown under our formalism, since each $\ETHmeasure{\hat{H}}{\Bqty*{\hat{A}}}$ provides a lower bound on $\ETHmeasure{\hat{H}}{\opSet}$.
In this sense, $\ETHmeasure{\hat{H}}{\opSet}$ is the most sensitive ETH measure.

\paragraph{Test observables.---}
We then introduce the set of test observables.
For $\particleNumber$-site spin-$S$ systems, we define the $m$-body operator space $\opSet_{\particleNumber}^{[0, m]}$ as the space of operators that can be expressed as a linear combination of operators acting nontrivially on at most $m$ spins~\cite{hamazaki2018atypicality}:
\begin{equation*}
    \opSet_{\particleNumber}^{[0, m]}
    \coloneqq \bigoplus_{m'=0}^{m} \vecspan\Bqty{ \hat{\sigma}_{x_{1}}^{(p_{1})} \cdots \hat{\sigma}_{x_{m'}}^{(p_{m'})} \mid \substack{ 1 \leq x_{1} < \cdots < x_{m'} \leq \sysSize \\ 0 < p_{j} < \dimLoc^2 } },
\end{equation*}
where $\Bqty*{ \hat{\sigma}_{x}^{(p)} }_{p=0}^{\dimLoc^2-1}$ is an orthonormal basis of the local operator space at site $x$ with $\hat{\sigma}_{x}^{(0)} = \hat{I}/\sqrt{\dimLoc}$ ($\hat{I}$ denotes the identity operator).
Here, $\opSet_{\particleNumber}^{[0,0]}$ should be understood as $\opSet_{\particleNumber}^{[0,0]} = \vecspan\!\Bqty*{\hat{I}}$.
For Bose and Fermi systems, we define $\opSet_{\particleNumber}^{[0, m]}$ to be the space of operators that can be written as a linear combination of products of $m$ annihilation operators and $m$ creation operators.
That is, we define $\opSet_{\particleNumber}^{[0, m]} \coloneqq \tilde{\opSet}_{\particleNumber}^{[0, m]} \cap \mathcal{L}(\mathcal{H})$ with 
\begin{equation*}
    \tilde{\opSet}_{\particleNumber}^{[0, m]} 
    \coloneqq \vecspan_{\mathbb{C}}\Bqty{ \hat{a}_{x_{1}}^{\dagger} \cdots \hat{a}_{x_{m}}^{\dagger} \hat{a}_{y_{1}} \cdots \hat{a}_{y_{m}} \mid \substack{ 1 \leq x_{j} \leq V \\ 1 \leq y_{j} \leq V } },
\end{equation*}
where $\hat{a}_{x}$ denotes the bosonic/fermionic annihilation operator at site $x$, and $\mathcal{L}(\mathcal{H})$ denotes the space of all Hermitian operators acting on the Hilbert space $\mathcal{H}$.
With these definitions, we have $\opSet_{\particleNumber}^{[0, m-1]} \subseteq \opSet_{\particleNumber}^{[0, m]}$ for any $m = 1,\cdots, \particleNumber$, which justifies the superscript $[0, m]$.

\paragraph{Motivating analysis in an idealized random-matrix model.---}
Before numerically testing Conjecture~\ref{conj_Conjecture} for systems with local and few-body interactions, we show the behavior of the ETH measure in the shell-wise locally rotationally invariant description of chaotic many-body eigenstates as used in Refs.~\cite{von2010proof, reimann2015generalization, Pappalardi_PhysRevLett2022z}. 
In this description, statistical properties of chaotic many-body eigenstates are modeled by a distribution that is invariant under local rotations (i.e., unitary transformations) within each narrow energy shell.
To be concrete, we divide the total Hilbert space into disjoint energy shells $\mathcal{H} = \bigoplus_j \mathcal{H}_j$, where $\mathcal{H}_j \coloneqq \vecspan\{ \ket*{E_{\alpha}} \colon \abs*{ E_{\alpha} - E_j } \leq \delta_j \}$ is an energy shell centered at energy $E_j$ with width $2 \delta_j$.
The shell width $\delta_j$ is chosen so that energy dependence within each shell is negligible while the shell still contains exponentially many eigenstates.
Let $\{ \ket*{j,\mu} \}_{\mu}$ be an orthonormal basis of $\mathcal{H}_j$.
After relabeling the energy eigenstates in $\mathcal{H}_j$ as $\{ \ket*{E_{(j,\mu)}} \}_{\mu}$, the two bases are related by a unitary
operator $\hat{U}^{(j)}$ acting on $\mathcal{H}_j$: $\ket*{E_{(j,\mu)}} = \hat{U}^{(j)}\ket*{j,\mu}$.
While $\hat{U}^{(j)}$ is deterministic once the reference basis $\{ \ket*{j,\mu} \}_{\mu}$ is fixed, it is modeled as a pseudorandom unitary whose distribution is invariant under any local rotation of the reference basis $\{ \ket*{j,\mu} \}_{\mu} \mapsto \{ \hat{V}^{(j)} \ket*{j,\mu} \}_{\mu}$.

For the test-observable spaces introduced in the previous section, the random-matrix calculation gives, with probability approaching one,
\begin{align}
    \ETHmeasure{\hat{H}}{ \opSet_{\particleNumber}^{[0,m]} }
    &\leq \mathbb{E}_{U}\!\bqty{ \sqrt{\dim \mathcal{H}}\, \ETHmeasure[2]{\hat{H}}{ \opSet_{\particleNumber}^{[0,m]} } } + \order{ \frac{ D_{\mathrm{min}}(E)^{\epsilon} }{ \sqrt{ D_{\mathrm{min}}(E) } } }
    \nonumber \\
    &\leq \sqrt{ \frac{ \dim\opSet_{\particleNumber}^{[0,m]} }{ D_{\mathrm{min}}(E) }} + \order{ \frac{ D_{\mathrm{min}}(E)^{\epsilon} }{ \sqrt{ D_{\mathrm{min}}(E) } } }
    \label{eq_RandomMatrixEstimate}
\end{align}
where $\epsilon \in (0,1/2)$ and $D_{\text{min}}(E) \coloneqq \displaystyle\min_{j\colon \mathcal{H}_j \cap \mathcal{H}_{E,\Delta\!{E}} \neq \emptyset} \dim\mathcal{H}_j$ is the minimum dimension of the energy shells overlapping the ETH test window $\mathcal{H}_{E,\Delta\!{E}}$.
Since $D_{\text{min}}(E) = e^{\particleNumber s_{\text{min}} + o(\particleNumber)} \ (s_{\text{min}} > 0)$ at finite temperature and
$\dim\opSet_{\particleNumber}^{[0,\alpha \particleNumber]}\leq e^{\particleNumber \Phi(\alpha) + o(\particleNumber)}$ with $\Phi(\alpha) \to 0$ as $\alpha \to 0$, this result predicts a change in the finite-size scaling behavior of the upper bound on $\ETHmeasure{\hat{H}}{ \opSet_{\particleNumber}^{[0,\alpha \particleNumber]} }$ at the point where $\Phi(\lowerBound^{\text{RM}}) = s_{\text{min}}$ and $\lowerBound^{\text{RM}} > 0$.
For $\alpha < \lowerBound^{\text{RM}}$, the upper bound in Eq.~\eqref{eq_RandomMatrixEstimate} vanishes exponentially in the thermodynamic limit.
Therefore, Conjecture~\ref{conj_Conjecture} holds true for this idealized random-matrix model.
The numerical analysis on systems with local and few-body interactions below uses the analytic $\particleNumber$-dependence in~\eqref{eq_RandomMatrixEstimate} to interpret the crossings of the computable bounds for local and few-body Hamiltonian ensembles.
The derivation of Eq.~\eqref{eq_RandomMatrixEstimate} is given in \Supple{}~\ref{SectionSM3}.

\paragraph{Numerical tests for locally interacting systems.---}
To test Conjecture~\ref{conj_Conjecture} for systems with local and few-body interactions, we introduce ensembles of such Hamiltonians for spin, Bose, and Fermi systems.
For spin systems, we consider the ensemble of Hamiltonians consisting of 2-local terms, i.e., those given by
\begin{equation}
    \hat{H} \coloneqq \sum_{\alpha,\beta = 0}^{3} J_{\alpha\beta} \qty( \sum_{j=1}^{\sysSize} \hat{\sigma}_{j}^{(\alpha)} \hat{\sigma}_{j+1}^{(\beta)} ),
    \label{eq_SpinHamiltonian}
\end{equation}
where $\hat{\sigma}^{(\alpha)}$ ($\alpha = 1, 2, 3$) are the Pauli operators, $\hat{\sigma}^{(0)} \coloneqq \hat{I}/\sqrt{2}$ with $\hat{I}$ being the identity operator, and the coefficients $\Bqty*{ J_{\alpha\beta} }$ are i.i.d. Gaussian random variables with zero mean and unit variance.
Since the Hamiltonian~\eqref{eq_SpinHamiltonian} has translation symmetry, we restrict ourselves to the zero-momentum sector in numerical calculations.
For Bose and Fermi systems, we consider the ensemble of Hamiltonians consisting of 3-local and 2-body terms, i.e., those given by $\hat{H} \coloneqq (\tilde{H} + \tilde{H}^{\dagger} + \hat{P} \tilde{H} \hat{P} + \hat{P} \tilde{H}^{\dagger} \hat{P} ) / 4$, where $\hat{P}$ is the reflection operator defined by $\hat{P} \hat{a}_{j} \hat{P} = \hat{a}_{\sysSize - j +1}$ with $\hat{a}_{j}$ being bosonic or fermionic annihilation operator acting on site $j$, and
\begin{equation}
    \tilde{H} \coloneqq \sum_{x_{1},x_{2},y_{1},y_{2} = 1}^{3} J_{x_{1}x_{2}}^{y_{1}y_{2}} \qty( \sum_{j=1}^{\sysSize} \hat{a}_{j+x_{1}}^{\dagger} \hat{a}_{j+x_{2}}^{\dagger} \hat{a}_{j+y_{1}} \hat{a}_{j+y_{2}} ).
    \label{eq_BosonFermionHamiltonian}
\end{equation}
Here, the coefficients $\Bqty*{ J_{x_{1}x_{2}}^{y_{1}y_{2}} }$ are i.i.d. complex Gaussian random variables satisfying $\mathbb{E}J_{x_{1}x_{2}}^{y_{1}y_{2}} = 0$, $\mathbb{E}\abs*{J_{x_{1}x_{2}}^{y_{1}y_{2}}}^2 = 1$, and $\mathbb{E}(J_{x_{1}x_{2}}^{y_{1}y_{2}})^2 = 0$.
Since the Hamiltonian $\hat{H}$ has translation and reflection symmetries, we restrict ourselves to the zero-momentum and even-parity sector in numerical calculations.
The filling factor $(n \coloneqq \particleNumber/\sysSize)$ is set to $n=1$ for Bose systems and $n=1/2$ for Fermi systems.

For these ensembles, we numerically evaluate the quantity $\Lambda_{2}^{[0,m]}$, which bounds the ETH measure from above and below as in Eq.~\eqref{eq_BoundsOnETHmeasure}.
(See the End Matter for details on the numerical calculations.)
The width $\var{E}$ of the microcanonical shell is set to $\var{E} = 0.2(E_{\mathrm{max}} - E_{\mathrm{min}})/\sysSize$, which is sufficiently small to suppress contributions from energy dependence for the system sizes considered.
Here, $E_{\mathrm{max/min}}$ denotes the maximum/minimum energy eigenvalue.
Figure~\ref{fig_Numerics_ShortRangeEnsemble} shows the averaged upper bound $\mathbb{E}_{\log}[\sqrt{\dim \mathcal{H}}\, \Lambda_{2}^{[0,m]}]$ of the (diagonal) ETH measure in the middle of the spectrum as a function of the ratio $m/\particleNumber$ for each ensemble.
Here, $\mathbb{E}_{\log}[X] \coloneqq \exp\!\qty( \frac{1}{k} \sum_{j=1}^{k} \log X_j)$ denotes the geometric mean.
We observe that the curves for different system sizes intersect approximately at a particular value, denoted by $\lowerBound$.
To the left of this crossing point~($m/\particleNumber \lesssim \lowerBound{}$), the averaged upper bound decreases with increasing system size, while it increases to the right ($m/\particleNumber \gtrsim \lowerBound{}$).
This $\particleNumber$-dependence qualitatively parallels that of the upper bound in Eq.~\eqref{eq_RandomMatrixEstimate} for an idealized random-matrix model at infinite temperature $D_{\mathrm{min}}(E) \sim \dim \mathcal{H}$, shown in the figure as dashed curves.
This qualitative agreement provides evidence that the ETH holds simultaneously for all $m$-body operators with $m/\particleNumber < \lowerBound{}$, yielding a concrete lower bound $\lowerBound{} \leq \alpha_{\ast}$ for the investigated systems and thereby supporting Conjecture~\ref{conj_Conjecture}.
It is noteworthy that such consistency is observed for systems with local and few-body interactions, even though Conjecture~\ref{conj_Conjecture} can be rigorously proved only for idealized (pseudo)random-matrix descriptions of many-body eigenstates.

Moreover, we extend our numerical analysis to varying normalized energies $x = (E - E_{\mathrm{min}})/(E_{\mathrm{max}} - E_{\mathrm{min}})$ and find that our conjecture holds true even for energy densities corresponding to finite temperatures.
That is, for values not far away from the center of the spectrum ($x \approx 0.5$), we observe behavior similar to that in Fig.~\ref{fig_Numerics_ShortRangeEnsemble}: 
the curves of the upper bound $\sqrt{\dim\mathcal{H}}\, \Lambda_{2}^{[0,m]}$ on the (diagonal) ETH measure for different system sizes intersect almost at a single value of $m/\particleNumber$, denoted by $\lowerBound(x)$ (see \Supple{}~\ref{SectionSM5}).
As shown in Fig.~\ref{fig_BoundsEnergyDependence}, this intersection point $\lowerBound(x)$ slightly decreases as the normalized energy deviates from the spectral center.
For small values of $x \lesssim 0.25$, the intersection is no longer observed, leaving open the possibility that $\alpha_{\ast} = 0$ in this low-energy region.
A similar crossing behavior is also observed for the lower bound $\Lambda_{2}^{[0,m]}$ of the ETH measure, with the intersection point denoted by $\upperBound(x)$.
For $\alpha > \upperBound(x)$, the lower bound remains $\order{1}$ in the thermodynamic limit, thereby implying the upper bound $\alpha_{\ast} \leq \upperBound(x)$.
As in the case for $\lowerBound(x)$, $\upperBound(x)$ decreases as the normalized energy decreases (see Fig.~\ref{fig_BoundsEnergyDependence}).
Note that the intersection points $\lowerBound(x)$ (or $\upperBound(x)$) are numerically identified as the value of $\alpha \coloneqq m/\particleNumber$ that minimizes the maximum difference between the bounds on the ETH measure, $\mathbb{E}_{\log}[\sqrt{\dim\mathcal{H}}\, \Lambda_{2}^{[0,\alpha\particleNumber]}]$ (or $\mathbb{E}_{\log}[\Lambda_{2}^{[0,\alpha\particleNumber]}]$), for different system sizes.

We also test Conjecture~\ref{conj_Conjecture} for concrete nonintegrable systems where the ETH is numerically verified to hold, namely, the Ising model with transverse and longitudinal fields~\cite{kim2014testing}, the Bose-Hubbard model at unit filling~\cite{biroli2010effect}, and spinless fermions with next-nearest-neighbor terms~\cite{rigol2009quantum} at half filling.
However, Conjecture~\ref{conj_Conjecture} is neither validated nor invalidated for these concrete models.
For these models, finite-size effects are so significant that we cannot safely determine whether the upper bound of the ETH measure decreases for large system sizes even for $m=1$~(see \Supple{}~\ref{SectionSM6}).
This indicates that the inequality $\seminorm{1}{\cdot} \leq \sqrt{\dim\mathcal{H}}\, \seminorm{2}{\cdot}$ is too loose to enable a conclusive test of Conjecture~\ref{conj_Conjecture} for those concrete physical models.
It remains to be an important future task to validate (or invalidate) Conjecture~\ref{conj_Conjecture} for prototypical nonintegrable systems, e.g., by finding a better upper bound on the ETH measure.

A complementary theoretical expectation also points to the need for an improved lower bound.
Given a plausible argument based on comparing the reduced density operators of energy eigenstates and thermal ensembles on subsystems~\cite{garrison2018does}, we believe that the ETH typically breaks down for $\opSet_{\particleNumber}^{[0,\particleNumber/2]}$, i.e., we expect $\alpha_{\ast} \leq 1/2$ both for realistic systems and idealized (pseudo)random-matrix models.
However, the numerical results for systems with local and few-body interactions~(Fig.~\ref{fig_BoundsEnergyDependence}) do not show $\upperBound{} < 0.5$.
This highlights an analytically challenging yet meaningful direction for future research---developing a tighter lower bound on the ETH measure by explicitly incorporating the locality and the few-body nature of observables.

Beyond quantum thermalization, the (semi-)norm $\seminorm{1}{\cdot}$ introduced in Eq.~\eqref{def_pseudoDistance1} serves more generally as a measure of closeness between quantum states relative to the operator set $\opSet$.
Consequently, this measure has potential utility in broader contexts beyond the ETH for comparing a time-evolving state with its steady state and assessing the operational accuracy of state approximations.
Moreover, it can provide a framework for constructing the space of macroscopic states from that of quantum states by identifying quantum states close to each other under the norm $\seminorm{1}{\cdot}$.
It remains an important future challenge to rigorously formulate the correspondence between microscopic and macroscopic states along this line and derive the macroscopic dynamics from the microscopic one.

\section{Data availability}
The numerical data plotted in the figures of this Article are available from the authors upon reasonable request.

\section{Code availability}
The source code used in the numerical calculations is available at Zenodo:~\url{https://doi.org/10.5281/zenodo.14709223}.

\section{Acknowledgments}
\begin{acknowledgments}
We are grateful to Takashi Mori for the discussions about the relation of our results to the previous ones.
S.S. is also grateful to Masaya Nakagawa for the valuable and helpful discussions on the dimension counting of the $m$-body space for Fermi systems.
This work was supported by KAKENHI Grant Numbers JP22H01152 from the Japan Society for the Promotion of Science (JSPS). 
S.S. was supported by KAKENHI Grant Number JP22J14935 from the Japan Society for the Promotion of Science (JSPS) and Forefront Physics and Mathematics Program to Drive Transformation (FoPM), a World-leading Innovative Graduate Study (WINGS) Program, the University of Tokyo.
R.H. was supported by JST ERATO-FS Grant Number JPMJER2204 and JST ERATO Grant Number JPMJER2302, Japan, and
JSPS KAKENHI Grant No. JP24K16982.
M.U. is supported by the CREST program \enquote{Quantum Frontiers} (Grant No.~JPMJCR2311) by the Japan Science and Technology Agency.
\end{acknowledgments}

\section{Competing interests}
The authors declare no competing interests.

\section{Author contributions}
All authors contributed to the manuscript writing, the discussion, and the interpretation of the analytical results and the numerical data.
S.S. conceived the project, carried out all the analytical and numerical calculations, and drafted and revised the manuscript.
R.H. regularly reviewed the analytical calculations, suggested improvements, and provided substantial feedback on the revision of the manuscript.
M.U. provided essential feedback on the clarity, structure, and overall presentation of the manuscript, and supervised the project.

\bibliography{reference}

\clearpage
\section{End Matter}
\paragraph{Scaling of the normalization constant $\opRange[q]{ \hat{A} }$.---}
Here, we demonstrate that the quantity $\hat{A} / \norm*{ \hat{A} }_{q}$ in the definition of $\seminorm{1}{\cdot}$ in Eq.~\eqref{def_pseudoDistance1} becomes thermodynamically intensive only when $q=\infty$.
To illustrate this, we analyze the $\particleNumber$-dependence of the normalization constant $\opRange[q]{ \hat{A} }$ for an extensive operator $\hat{M}_{z} \coloneqq \sum_{j=1}^{\particleNumber} \hat{\sigma}^{(z)}_{j}$, where $\hat{\sigma}^{(z)}$ is the Pauli $z$-operator.
Let $\ket{\uparrow}$ and $\ket{\downarrow}$ denote its eigenstates with eigenvalues $+1$ and $-1$, respectively.
The eigenstates of $\hat{M}_{z}$ are tensor products of $\ket{\uparrow}$ and $\ket{\downarrow}$, and their eigenvalues are $m_{j} \coloneqq -\particleNumber+2j \ (j=0,\cdots,\particleNumber)$, where $j$ is the number of up spins.
Thus, we have $\norm*{\hat{M}_{z}}_{\infty} = N$, and $\hat{M}_{z} / \norm*{\hat{M}_{z}}_{\infty}$ is intensive by construction.
On the other hand, for any finite $q\, (\geq 1)$, we have
\begin{align}
    \opRange[q]{ \hat{M}_{z} }^{q} &= 
    \sum_{j=0}^{\particleNumber} \binom{\particleNumber}{j} \abs*{-\particleNumber+2j}^{q}.
\end{align}
This sum is equal to the $q$th absolute moment of the sum of $\particleNumber$ independent Rademacher random variables $\Bqty*{ \xi_{k} }_{k=1}^{\particleNumber}$, where $\Prob[\xi_{k} = \pm 1] = 1/2$:
\begin{equation}
    \opRange[q]{ \hat{M}_{z} }^{q} = 2^{\particleNumber} \mathbb{E}[\abs*{S_{\particleNumber}}^q]\qq{where} 
    S_{N} \coloneqq \sum_{k=1}^{\particleNumber} \xi_{k}.
\end{equation}
Indeed, if exactly $j$ of the $\Bqty{\xi_{k}}$ equal $+1$, then $S_{\particleNumber}=-\particleNumber+2j$, which occurs with probability $2^{-\particleNumber} \binom{\particleNumber}{j}$.
The mean and variance of $S_{\particleNumber}$ are $\mathbb{E}[S_{\particleNumber}] = 0$ and $\mathbb{V}[S_{\particleNumber}]=\particleNumber$, respectively.
Thus, the central limit theorem implies that the normalized variable $Z_{\particleNumber} \coloneqq S_{\particleNumber} / \sqrt{\particleNumber}$ converges in distribution to a standard normal variable.
As a result, we obtain
\begin{equation}
    \opRange[q]{ \hat{M}_{z} }^{q} = 2^{\particleNumber} \particleNumber^{\frac{q}{2}}\, \bqty{ \frac{ 2^{q/2} }{\sqrt{\pi}} \Gamma\qty(\frac{q+1}{2}) +\order{\particleNumber^{-1}} }.
\end{equation}
Thus, for any fixed $q$ and sufficiently large $\particleNumber$, the normalization constant $\opRange[q]{ \hat{M}_{z} }$ scales as $\sim 2^{\particleNumber / q} \sqrt{\particleNumber}$.
Consequently, $\hat{M}_{z}/\norm*{ \hat{M}_{z} }_{q}$ becomes thermodynamically intensive only when $q=\infty$.

\paragraph{Proof of Proposition~\ref{Proposition1}.---}
The inequality~\eqref{eq_BoundsOn1norm} follows from the simple inequality between the operator norm and the Hilbert--Schmidt norm:
\begin{equation}
    \frac{1}{\sqrt{\rank \hat{A}}} \norm*{ \hat{A} }_{2} \leq \norm*{ \hat{A} }_{\infty} \leq \norm*{ \hat{A} }_{2}.
\end{equation}
Substituting this inequality into the definition~\eqref{def_pseudoDistance1} gives the inequality~\eqref{eq_BoundsOn1norm}.

To derive the formula~\eqref{eq_ComputableFormulaFor2Norm}, we use the expansion of $\hat{A} \in \opSet + \mathbb{R} \hat{I}$ as $\hat{A} = \sum_{\ell} c_{\ell} \hat{\Lambda}_{\ell}$, where $c_{\ell} = \tr(\hat{\Lambda}_{\ell}^{\dagger} \hat{A}) \in \mathbb{R}$.
Substituting this expansion in Eq.~\eqref{def_pseudoDistance1} gives
\begin{align}
    \seminorm*{2}{\hat{X}}
    %
    &= \sup_{ \substack{ \vec{c} \in \mathbb{R}^{M} \colon \vec{c} \neq 0}} 
        \abs{ \sum_{\ell} \frac{ c_{\ell} }{ \norm*{\vec{c}}_{2} } \tr\qty( \hat{\Lambda}_{\ell}^{\dagger} \,\hat{X} )  }
    \nonumber \\
    &= \sup_{ \substack{ \vec{c} \in \mathbb{R}^{M} \colon \norm*{\vec{c}}_{2} = 1}} 
        \abs{ \vec{c} \cdot \qty( \Re\vec{X} + i \Im\vec{X} )  }
    \nonumber \\
    &= \sup_{ \substack{ \vec{c} \in \mathbb{R}^{M} \colon \norm*{\vec{c}}_{2} = 1}} 
        \sqrt{ \vec{c}^{T} \mathcal{Q} \vec{c} },
\end{align}
where $\mathcal{Q} \coloneqq (\Re\vec{X})(\Re\vec{X})^{T} + (\Im\vec{X})(\Im\vec{X})^{T}$ is a symmetric matrix.
Since the rank of $\mathcal{Q}$ is at most 2, it is straightforward to compute the eigenvalues of $\mathcal{Q}$, and we obtain the formula~\eqref{eq_ComputableFormulaFor2Norm}.
\qed{}

\bigskip
\paragraph{Details on the numerical calculation.---}
In our numerical tests of Conjecture~\ref{conj_Conjecture} for systems with local and few-body interactions, we calculate the quantity
\begin{equation}
    \Lambda_{2}^{[0,m]}
    \coloneqq 
    \max_{ \ket*{E_{\alpha}} \in \mathcal{H}_{E_{x}, \Delta\!{E}} } \seminorm[\opSet_{\particleNumber}^{[0,m]}]{2}{ \dyad*{ E_{\alpha} }{ E_{\alpha} } - \MCE[\var{E}](E_{\alpha}) },
    \label{eqS_quasiETHmeasure}
\end{equation}
which provides a lower bound on the ETH measure and an upper bound on the diagonal ETH measure.
Here, the energy window $\mathcal{H}_{E_{x}, \Delta\!{E}}$ for testing the ETH is centered at $E_{x} \coloneqq x E_{\mathrm{max}} + (1-x) E_{\mathrm{min}}$ with width $2\Delta{E} \coloneqq 0.05 (E_{\mathrm{max}} - E_{\mathrm{min}})$, where $E_{\mathrm{max}}$ and $E_{\mathrm{min}}$ are the maximum and minimum eigenvalues of $\hat{H}$, respectively. 
Since the spectral width $(E_{\mathrm{max}} - E_{\mathrm{min}})$ is extensive for locally interacting systems~\cite{Tasaki2018-bz}, the normalized energy $x = (E - E_{\mathrm{min}})/(E_{\mathrm{max}} - E_{\mathrm{min}})$ represents the energy density up to linear rescaling.
The width of the microcanonical energy shell is set to be $\var{E} = 0.2 (E_{\mathrm{max}} - E_{\mathrm{min}}) / V$, which we confirm is sufficiently small to exclude contributions to $\Lambda_{2}^{[0,m]}$ from the energy dependence of the microcanonical average for the system sizes considered (up to 18 spins for spin systems, and up to $11$ particles for Bose and Fermi systems at unit and half filling, respectively).

The quantity $\Lambda_{2}^{[0,m]}$ is computed using the explicit formula in Proposition~\ref{Proposition1}.
The calculation of the 2-norm for a given $\hat{X}$ costs $\order{M D}$ steps, assuming that an orthogonal basis of $\opSet_{\particleNumber}^{[0,m]}$ consists of sparse matrices in the Fock basis, where $M \coloneqq \dim(\opSet_{\particleNumber}^{[0,m]} + \mathbb{R} \hat{I})$ and $D \coloneqq \dim \mathcal{H}$.
Thus, the computation of the bounds including off-diagonal elements costs $\order{M D^3}$ steps, which is much heavier than the cost $\order{M D^2}$ for the diagonal elements and the cost $\order{D^3}$ of numerical diagonalization of $\hat{H}$.
Therefore, we restrict ourselves to diagonal elements, as in Eq.~\eqref{eqS_quasiETHmeasure}.

The formula in Proposition~\ref{Proposition1} requires an orthonormal basis $\Bqty*{ \hat{\Lambda}_{\mu} }_{\mu=1}^{M}$ of $\opSet_{\particleNumber}^{[0,m]} + \mathbb{R} \hat{I}$.
In spin-1/2 systems, such a basis is given by
\begin{equation*}
    \Bqty*{ \hat{\Lambda}_{\mu} } = 2^{-\frac{\sysSize}{2}} \Bqty{ \hat{\sigma}^{\vb{p}} \coloneqq \hat{\sigma}_{1}^{(p_{1})} \hat{\sigma}_{2}^{(p_{2})} \cdots \hat{\sigma}_{\sysSize}^{(p_{\sysSize})} \mid m(\vb{p}) \leq m },
\end{equation*}
where $\vb{p} \coloneqq (p_{1}, p_{2}, \cdots, p_{\sysSize}) \in \Bqty*{0,1,2,3}^{\!\!\sysSize{}}$ with $\hat{\sigma}^{(0)} = \hat{I}$~(identity), and $m(\vb{p})$ denotes the number of nonzero components in $\vb{p}$.
For Bose and Fermi systems, the basis operators $\Bqty*{ \hat{\boldsymbol{\alpha}}(\vb{x},\vb{y}) \coloneqq \hat{a}_{x_{1}}^{\dagger} \cdots \hat{a}_{x_{m}}^{\dagger} \hat{a}_{y_{1}} \cdots \hat{a}_{y_{m}} }$ are not mutually orthogonal under the Hilbert–Schmidt inner product.
Here, $\vb{x} = (x_{1},\cdots,x_{m})$ and $\vb{y} = (y_{1},\cdots,y_{m})$, and $\hat{a}$ denotes either a bosonic or fermionic annihilation operator.
To obtain an orthonormal basis, we construct the Gram matrix $\vb{G}$ given by $G_{(\vb{x},\vb{y}), (\vb{x}',\vb{y}')} \coloneqq \tr_{N}\!\bqty{ \hat{\boldsymbol{\alpha}}(\vb{x},\vb{y})^{\dagger} \hat{\boldsymbol{\alpha}}(\vb{x}',\vb{y}') }$, where $\tr_{N}$ denotes the trace over the $N$-particle Hilbert space.
Since $\vb{G}$ is Hermitian and positive definite, the Cholesky decomposition yields $\vb{G} = \vb{R}^{\dagger} \vb{R}$, where $\vb{R}$ is an upper triangular matrix.
An orthonormal basis is then constructed as $\hat{\Lambda}_{\nu} \coloneqq \sum_{\mu} \hat{\boldsymbol{\alpha}}_{\mu} (\vb{R}^{-1})_{\mu\nu}$.
Thus, the coefficient vector $\vec{X}$ appearing in the formula in Proposition~\ref{Proposition1} is obtained as
\begin{align}
    \vec{X} = (\vb{R}^{\dagger})^{-1} \vec{v},\qq{where} v_{\mu} \coloneqq \tr[ \hat{\boldsymbol{\alpha}}_{\mu}^{\dagger} \hat{X} ].
    \label{eqS_2normComputableFormula2}
\end{align}
We employ this expression to calculate the 2-norm for Bose and Fermi systems.

In general, the evaluation of the formula~\eqref{eqS_2normComputableFormula2} costs $\order{D M^2}$ operations to construct the Gram matrix $\vb{G}$, and $\order{M^3}$ operations to compute its Cholesky factor $\vb{R}$.
These costs readily exceed the diagonalization cost of the Hamiltonian, since the operator-space dimension $M$ can be much larger than the Hilbert-space dimension $D$.
However, we find that the Gram matrix $\vb{G}$ has a block-diagonal structure and that the size of each block is reasonably small.
Moreover, for translationally invariant $\hat{X}$ satisfying $\hat{T} \hat{X} \hat{T}^{\dagger} = \hat{X}$, with $\hat{T}\hat{a}_{j} \hat{T}^{\dagger} \coloneqq \hat{a}_{j+1}$, these blocks can be grouped into orbits under the action of the shift operator $\hat{T}$.
Since all blocks in the same orbit contribute identically to $\seminorm*{2}{\hat{X}}$, it suffices to compute a single representative per orbit, further reducing the computational cost by a factor of $1/\sysSize$.

The block-diagonal structure of the Gram matrix $\vb{G}$ arises from the vanishing of many inner products $\tr_{N}[\hat{\boldsymbol{\alpha}}(\vb{x},\vb{y})^{\dagger} \hat{\boldsymbol{\alpha}}(\vb{x}',\vb{y}')]$.
This inner product is nonzero only if there exists a basis state $\ket*{\vb{n}} \coloneqq (\hat{a}_{1}^{\dagger})^{n_{1}} \cdots (\hat{a}_{\sysSize}^{\dagger})^{n_{\sysSize}} \ket{0}$ such that $\hat{\boldsymbol{\alpha}}(\vb{x},\vb{y}) \ket*{\vb{n}} = C_{\vb{n}}\, \hat{\boldsymbol{\alpha}}(\vb{x}',\vb{y}') \ket*{\vb{n}}$ for some $C_{\vb{n}} \neq 0$.
This is because each $\hat{\boldsymbol{\alpha}}$ maps a Fock state to another Fock state.
By tracking the net increase and decrease in occupation numbers at each site, this requirement leads to a simple condition involving the multisets $\mathcal{X} \coloneqq \Bqty*{ x_{j} }_{j=1}^{m}$ and $\mathcal{Y} \coloneqq \Bqty*{ y_{j} }_{j=1}^{m}$, and similarly $\mathcal{X}'$ and $\mathcal{Y}'$:
\begin{equation}
    \mathcal{X}\setminus \mathcal{Y} = \mathcal{X}'\setminus \mathcal{Y}'
    \qq{and}
    \mathcal{Y}\setminus \mathcal{X} = \mathcal{Y}'\setminus \mathcal{X}'.
\end{equation}
This condition partitions the operator basis $\Bqty*{\hat{\boldsymbol{\alpha}}(\vb{x},\vb{y})}$ into disjoint subsets within which the Gram matrix may have nonzero elements, resulting in its block-diagonal structure.
As an illustrative example, we show the block structure of the Gram matrix in two cases with $m = 4$ and $\particleNumber = 11$.
For Bose systems at unit filling, i.e., $\sysSize = \particleNumber = 11$, the total operator-space dimension is $M = \num{1002001}$, and the Gram matrix has the following block structure:
\[
\begin{tabular}{|r|r|r|} \hline
  \text{Block size} &
  \multicolumn{1}{c|}{\shortstack{\strut{} \# of blocks \\ of the same size}} &
  \multicolumn{1}{c|}{\shortstack{\strut{} \# of orbits \\ by the shift operation}} \\ \hline
  \num{1001} \hspace{1ex} & \num{1}      \hspace{1ex} & \num{1}     \hspace{4ex}  \\ 
   \num{286} \hspace{1ex} & \num{110}    \hspace{1ex} & \num{10}    \hspace{4ex} \\ 
    \num{66} \hspace{1ex} & \num{3080}   \hspace{1ex} & \num{280}   \hspace{4ex} \\ 
    \num{11} \hspace{1ex} & \num{40370}  \hspace{1ex} & \num{3670}  \hspace{4ex} \\ 
     \num{1} \hspace{1ex} & \num{322190} \hspace{1ex} & \num{29290} \hspace{4ex} \\ \hline
\end{tabular}
\]
For spinless Fermi systems at half filling, i.e., $\sysSize = 2 \particleNumber = 22$, the dimension becomes $M = \num{53509225}$, and the corresponding block structure is:
\[
\begin{tabular}{|r|r|r|} \hline
  \text{Block size} &
  \multicolumn{1}{c|}{\shortstack{\strut{} \# of blocks \\ of the same size}} &
  \multicolumn{1}{c|}{\shortstack{\strut{} \# of orbits \\ by the shift operation}} \\ \hline
  \num{7315} \hspace{1ex} &   \num{1}       \hspace{1ex} &  \num{1}       \hspace{4ex} \\ 
  \num{1140} \hspace{1ex} &  \num{462}      \hspace{1ex} &  \num{21}      \hspace{4ex} \\ 
   \num{153} \hspace{1ex} &  \num{43890}    \hspace{1ex} &  \num{2000}    \hspace{4ex} \\ 
    \num{16} \hspace{1ex} &  \num{1492260}  \hspace{1ex} &  \num{67830}   \hspace{4ex} \\ 
     \num{1} \hspace{1ex} &  \num{22383900} \hspace{1ex} &  \num{1017540} \hspace{4ex} \\ \hline
\end{tabular}
\]
In both cases, the vast majority of blocks are extremely small compared with the total operator-space dimension $M$.
For example, more than 90\% of them have size at most $16$.
This structure of the Gram matrix enables the efficient blockwise evaluation of the 2-norm, making it feasible to test Conjecture~\ref{conj_Conjecture} numerically for Bose and spinless Fermi systems up to $\particleNumber = 11$ particles.

\addtocontents{toc}{\string\tocdepth@restore} 
\clearpage\clearpage
\makeatletter
   	\c@secnumdepth=4
    \def\@pointsize{11}
	\expandafter\@process@pointsize\expandafter{\@pointsize@default}%
	\appdef\setup@hook{\normalsize}%
	\setup@hook
\makeatother
\setcounter{equation}{0}
\setcounter{figure}{0}
\setcounter{section}{0}
\setcounter{table}{0}
\renewcommand{\theequation}{S\arabic{equation}}
\renewcommand{\thefigure}{S\arabic{figure}}
\renewcommand{\theHequation}{\theequation}
\renewcommand{\theHfigure}{\thefigure}

\renewcommand{\baselinestretch}{1.10}
\normalsize
\title{
    \Large Supplementary Material: Bounds on eigenstate thermalization
}
\date{\today}
\maketitle
\onecolumngrid
\tableofcontents

\newcommand{\sectionS}[1]{%
    \refstepcounter{section}%
    \phantomsection%
    \begingroup
      \def\sectionStitle{#1}%
      \patchcmd{\sectionStitle}{\protect\\}{ }{}{}%
      \patchcmd{\sectionStitle}{\\}{ }{}{}%
      \addcontentsline{toc}{section}{%
        \protect\numberline{\thesection}%
        \sectionStitle
      }%
    \endgroup
    \begin{center}
      \normalfont\small\bfseries
      \thesection.\quad \MakeUppercase{#1}
    \end{center}%
}

\clearpage

\sectionS{Seminorm $\seminorm{1}{\cdot}$ as a unified measure \protect\\ of quantum-thermal equilibrium}
\label{SectionSM1}
As mentioned in the main text, the seminorm $\seminorm{1}{\cdot}$ with an appropriate choice of $\mathcal{A}$ serves as a unified measure for a rich variety of notions about quantum thermal equilibrium.
In this section, we provide some examples.

\subsection{Subsystem thermal equilibrium}
For any subsystem $\subSystem$ and any $\hat{A}_{\subSystem} \otimes \mathrm{id}_{\subSystem^{c}} \in \mathcal{L}(\mathcal{H}_{\subSystem}) \otimes \mathrm{id}_{\subSystem^{c}}$, we have $\opRange[\infty]{ \hat{A}_{\subSystem} \otimes \mathrm{id}_{\subSystem^{c}} } = \opRange[\infty]{ \hat{A}_{\subSystem} }$.
Here, $\mathcal{L}(\mathcal{H}_{\subSystem})$ denotes the space of all Hermitian operators acting on the Hilbert space $\mathcal{H}_{\subSystem}$.
By setting $\opSet = \mathcal{L}(\mathcal{H}_{\subSystem}) \otimes \mathrm{id}_{\subSystem^{c}}$, we have
\begin{align}
    \seminorm*{1}{\hat{X}}
    = \max_{ \hat{A} \in \mathcal{L}(\mathcal{H}_{\subSystem}) \otimes \mathrm{id}_{\subSystem^{c}} } \abs{ \tr\qty( \frac{ \hat{A} }{ \opRange[\infty]{ \hat{A} } } \hat{X} ) } 
    %
    = \max_{ \hat{A}_{\subSystem} \in \mathcal{L}(\mathcal{H}_{\subSystem}) } \abs{ \tr\qty( \frac{ \hat{A}_{\subSystem} }{ \opRange[\infty]{ \hat{A}_{\subSystem} } } \tr_{\subSystem^{c}}( \hat{X} ) ) } 
    = \norm{ \tr_{\subSystem^{c}}( \hat{X} ) }_{1}.
\end{align}
Thus, the seminorm $\seminorm{1}{\cdot}$ reduces to the trace norm on a subsystem $\subSystem$ for $\opSet = \mathcal{L}(\mathcal{H}_{\subSystem}) \otimes \mathrm{id}_{\subSystem^{c}}$, and the smallness of $\seminorm{1}{ \hat{\sigma} - \hat{\rho}_{\mathrm{th}} }$ for a thermal ensemble $\hat{\rho}_{\mathrm{th}}$ ensures that quantum states $\hat{\sigma}$ are in thermal equilibrium in a subsystem $\subSystem$.

\subsection{Microscopic thermal equilibrium~(MITE)~\citeSM{Goldstein2015SM,Goldstein2017SM,mori2018thermalizationSM}}
The notion of microscopic thermal equilibrium~(MITE) is introduced in Refs.~\citeSM{Goldstein2015SM, Goldstein2017SM} as follows.
\begin{framedDefinition}{Microscopic thermal equilibrium~(MITE)~\citeSM{Goldstein2017SM}}{}
    A state $\hat{\sigma}$ is said to be in microscopic thermal equilibrium~(MITE) on a length scale $l_{0}$ if it satisfies
    \begin{align}
        \norm{ \tr_{\subSystem^{c}}(\hat{\sigma}) - \tr_{\subSystem^{c}}(\hat{\rho}_{\mathrm{th}})  }_{1} < \epsilon
    \end{align}
    for every subsystem $\subSystem$ with $\mathop{\mathrm{Diam}}(\subSystem) \leq l_{0}$, where $\epsilon\ll 1$
    and $\mathop{\mathrm{Diam}}(\subSystem) \coloneqq \sup_{ x,y\in \subSystem } d(x,y)$ defines the diameter of $\subSystem$ for a given distance measure $d$.
\end{framedDefinition}
As mentioned in Ref.~\citeSM{Goldstein2017SM}, MITE can be regarded as the thermal equilibrium relative to
\begin{equation}
    \opSet_{\mathrm{MITE}} = \bigcup_{\subSystem\colon \mathop{\mathrm{Diam}}(\subSystem) \leq l_{0}} \mathcal{L}(\mathcal{H}_{\subSystem}) \otimes \mathrm{id}_{\subSystem^{c}}.
\end{equation}
Then, we have the following proposition.
\begin{framedProposition}{}{}
    \label{prop_RelationToMITE}
    Let $\hat{\sigma}$ be an arbitrary quantum state and $\hat{\rho}_{\mathrm{th}}$ be a thermal ensemble.
    Then, 
    \begin{equation}
        \text{$\hat{\sigma}$ is in MITE.} 
        \iff 
        \seminorm[ \opSet_{\mathrm{MITE}} ]{1}{ \hat{\sigma} - \hat{\rho}_{\mathrm{th}} }
        \leq \epsilon,
    \end{equation}
    where $\epsilon\ll 1$.
\end{framedProposition}
\begin{proof}
    The proof goes as follows:
    \begin{align}
        \seminorm[ \opSet_{\mathrm{MITE}} ]{1}{ \hat{\sigma} - \hat{\rho}_{\mathrm{th}} }
        &= \max_{ \hat{A} \in \opSet_{\mathrm{MITE}} } \abs{ \tr\qty( \frac{ \hat{A} }{ \opRange[\infty]{ \hat{A} } } (\hat{\sigma} - \hat{\rho}_{\mathrm{th}}) ) } \nonumber \\
        &= \max_{\subSystem\colon \mathop{\mathrm{Diam}}(\subSystem) \leq l_{0}} 
        \max_{ \hat{A}_{S} \in \mathcal{L}(\mathcal{H}_{S}) } \abs{ \tr_{S}\qty( \frac{ \hat{A}_{S} }{ \opRange[\infty]{ \hat{A}_{S} } } \qty\Big( \tr_{S^{c}}(\hat{\sigma}) - \tr_{S^{c}}(\hat{\rho}_{\mathrm{th}}) ) ) } \nonumber \\
        &= \max_{\subSystem\colon \mathop{\mathrm{Diam}}(\subSystem) \leq l_{0}} 
        \norm{ \tr_{S^{c}}(\hat{\sigma}) - \tr_{S^{c}}(\hat{\rho}_{\mathrm{th}}) }_{1}.
    \end{align}
\end{proof}

Mori et al.~\citeSM{mori2018thermalizationSM} extend the notion of MITE by replacing the spatial constraint $\mathrm{Diam}(\subSystem) \leq l_{0}$ with a \enquote{few-body} constraint as $\abs*{\subSystem} = k$ with an integer $k$ of $\order{1}$, i.e., they consider 
\begin{equation}
    \opSet_{\mathrm{MITE}}^{(\mathrm{few})} = \bigcup_{\subSystem\colon \abs*{\subSystem} = k} \mathcal{L}(\mathcal{H}_{\subSystem}) \otimes \mathrm{id}_{\subSystem^{c}}
\end{equation}
in addition to $\opSet_{\mathrm{MITE}}$.
The same proof for Proposition~\ref{Proposition1} applies to the MITE with respect to $\opSet_{\mathrm{MITE}}^{(\mathrm{few})}$, and we have the following proposition:
\begin{framedProposition}{}{}
    Let $\hat{\sigma}$ be an arbitrary quantum state and $\hat{\rho}_{\mathrm{th}}$ be a thermal ensemble.
    Then, 
    \begin{equation}
        \text{$\hat{\sigma}$ is in MITE with respect to $\opSet_{\mathrm{MITE}}^{(\mathrm{few})}$} 
        \iff 
        \seminorm[\opSet_{\mathrm{MITE}}^{(\mathrm{few})}]{1}{ \hat{\sigma} - \hat{\rho}_{\mathrm{th}}} 
        \leq \epsilon,
    \end{equation}
    where $\epsilon\ll 1$.
\end{framedProposition}

\subsection{Macroscopic thermal equilibrium~(MATE)~\citeSM{Goldstein2015SM,Goldstein2017SM,mori2018thermalizationSM}}
The notion of macroscopic thermal equilibrium~(MATE) is introduced in Refs.~\citeSM{Goldstein2015SM, Goldstein2017SM} as follows.
\begin{framedDefinition}{Macroscopic thermal equilibrium~(MATE)~\citeSM{Goldstein2017SM}}{}
    We consider a collection of macro observables $\hat{M}_{1},\cdots, \hat{M}_{K}$.
    By suitably coarse-graining the operators $\hat{M}_{1},\cdots, \hat{M}_{K}$, 
    we expect to obtain a set of mutually commuting operators $\tilde{M}_{1},\cdots, \tilde{M}_{K}$ with $\tilde{M}_{j} \approx \hat{M}_{j}$ for all $j=1,\cdots, K$.
    We take $\tilde{M}_{1}$ as the coarse-grained Hamiltonian, whose eigenspaces are energy shells $\mathcal{H}_{E,\Delta{E}}$.

    Since $\tilde{M}_{1},\cdots, \tilde{M}_{K}$ commute with each other, we can simultaneously diagonalize them, and the energy shell $\mathcal{H}_{E,\Delta{E}}$ can be decomposed accordingly as $\mathcal{H}_{E,\Delta{E}} = \bigoplus_{\nu} \mathcal{H}_{\nu}$.
    Here, $\mathcal{H}_{\nu}$ is called macro-spaces, and we denote the projector onto $\mathcal{H}_{\nu}$ by $\hat{\Pi}_{\nu}$.

    In each $\mathcal{H}_{E,\Delta{E}}$, it is expected that one macro-space called thermal equilibrium macro-space $\mathcal{H}_{\mathrm{eq}}$ covers the most of the dimensions of $\mathcal{H}_{E,\Delta{E}}$, i.e.,
    \begin{equation}
        \frac{ \dim \mathcal{H}_{\mathrm{eq}} }{ \dim \mathcal{H}_{E,\Delta{E}} } = 1 - \tilde{\epsilon},
    \end{equation}
    where $\tilde{\epsilon} \ll 1$.
    Without loss of generality, we set $\mathcal{H}_{\mathrm{eq}} = \mathcal{H}_{\nu=1}$.

    Under these setups, a state $\hat{\sigma}$ supported on $\mathcal{H}_{E,\Delta E}$ is said to be in macroscopic thermal equilibrium~(MATE) if and only if
    \begin{equation}
        \tr( \hat{\sigma} \hat{\Pi}_{\mathrm{eq}} ) \geq 1 - \delta
    \end{equation}
    for a sufficiently small tolerance $\delta > 0$.
\end{framedDefinition}

As mentioned in Ref.~\citeSM{Goldstein2017SM}, MATE can be regarded as the thermal equilibrium relative to the (coarse-grained) macroscopic observables $\tilde{M}_{1},\cdots, \tilde{M}_{K}$.
Because we focus on the joint distribution of the observed values of $\tilde{M}_{1},\cdots, \tilde{M}_{K}$ in MATE, $\opSet$ for MATE is given by
\begin{equation}
    \opSet_{\mathrm{MATE}} = \Bqty*{ \hat{\Pi}_{\mathrm{eq}} }.
    \label{eq:OpSetForMATE}
\end{equation}
Then, we have the following proposition.
\begin{framedProposition}{}{}
    Let $\hat{\sigma}$ be an arbitrary quantum state.
    Then, 
    \begin{equation}
        \text{$\hat{\sigma}$ is in MATE with tolerance $\delta \, (\geq 2\tilde{\epsilon})$}.
        \iff 
        \seminorm[\opSet_{\mathrm{MATE}}]{1}{ \hat{\sigma}- \MCE[\delta E] } \leq \delta - \tilde{\epsilon}.
    \end{equation}
\end{framedProposition}
\begin{proof}
    For $\opSet_{\mathrm{MATE}} = \Bqty*{ \hat{\Pi}_{\mathrm{eq}} }$, we have
    \begin{align}
        \seminorm[\opSet_{\mathrm{MATE}}]{1}{ \hat{\sigma}- \MCE[\delta E] } 
        &= \abs{ \tr( \hat{\sigma} \hat{\Pi}_{\mathrm{eq}} ) - \frac{ \dim \mathcal{H}_{\mathrm{eq}} }{ \dim \mathcal{H}_{E,\Delta{E}} } }.
    \end{align}
    Therefore, inequality $\seminorm[\opSet_{\mathrm{MATE}}]{1}{ \hat{\sigma}- \MCE[\delta E] }  \leq \epsilon$ is equivalent to
    \begin{equation}
        1-(\tilde{\epsilon}+\epsilon) 
        \leq \tr( \hat{\sigma} \hat{\Pi}_{\mathrm{eq}} )
        \leq 1+(\epsilon-\tilde{\epsilon}).
    \end{equation}
    Setting $\epsilon \coloneqq \delta - \tilde{\epsilon}\ (\geq \tilde{\epsilon})$, we obtain
    \begin{equation}
        \seminorm[\opSet_{\mathrm{MATE}}]{1}{ \hat{\sigma}- \MCE[\delta E] }
        \leq \delta - \tilde{\epsilon}
        \iff 1 - \delta \leq \tr( \hat{\sigma} \hat{\Pi}_{\mathrm{eq}} ),
    \end{equation}
    which is the desired result.
\end{proof}

Apart from quantum thermalization, the (semi-)norm $\seminorm{1}{\cdot}$ introduced in Eq.~\eqref{def_pseudoDistance1} in the main text serves as a measure of the closeness between two quantum states $\hat{\sigma}$ and $\hat{\rho}$ relative to $\opSet$.
Thus, it can be used in various situations other than quantum thermal equilibrium and the ETH, such as a comparison between the state during time evolution and the steady state.
In particular, it can be used to construct the space of macroscopic states from that of quantum states by identifying quantum states that are very close to each other in terms of $\seminorm{1}{\cdot}$.
It merits further study to formulate the correspondence between microscopic and macroscopic states in this direction, thereby deriving the macroscopic dynamics from the microscopic one.

\begin{figure}[h]
    \centering
    \includegraphics[width=\linewidth]{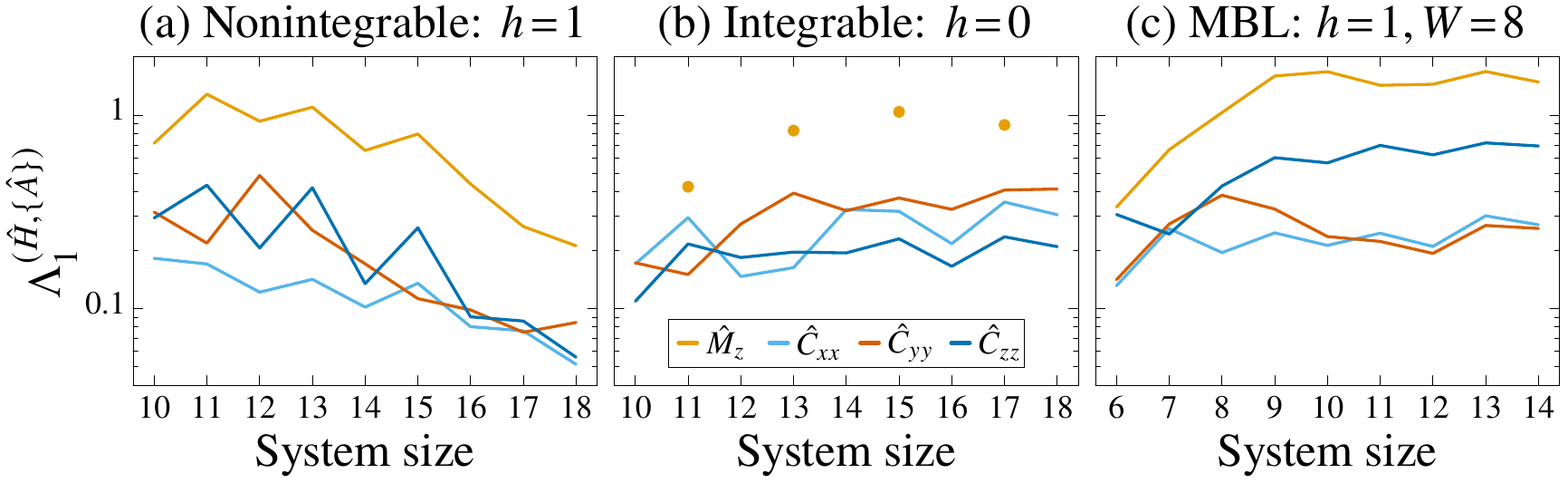}
    \caption{\textbf{ETH measure for representative observables in the XYZ model.}
    The ETH measure is plotted against the system size in the
        (a) nonintegrable regime (periodic boundary, $h = 1$, $W=0$), 
        (b) integrable regime (periodic boundary, $h = 0$, $W=0$), and 
        (c) many-body localized regime (open boundary, $h = 1$, $W = 8$).
        For panel~(b), the data for $\hat{M}_{z}$ are shown only at odd system sizes, because they vanish for even sizes, where the energy eigenstates are eigenstates of the $\pi$-rotation operator around the $x$ axis.
        The exponential decay in (a) and the saturation in (b) and (c) demonstrate that our ETH measure correctly detects both the validity and breakdown of the ETH.}
    \label{fig_XYZ_ETHmeasure_vsN}
\end{figure}

\clearpage
\sectionS{Validity of the ETH measure $\ETHmeasure{\hat{H}}{\Bqty*{\hat{A}}}$}
\label{SectionSM2}
To demonstrate the validity of our ETH measure $\ETHmeasure{\hat{H}}{\opSet}$ for the case of a single observable, we compute $\ETHmeasure{\hat{H}}{\Bqty*{\hat{A}}}$ for representative one- and two-body observables in the XYZ spin chain whose Hamiltonian is given by
\begin{equation}
    \hat{H} = \sum_{j=1}^{\sysSize} \sum_{\alpha=1}^{3} J_{\alpha} \hat{\sigma}_{j}^{(\alpha)} \hat{\sigma}_{j+1}^{(\alpha)} + \sum_{j=1}^{\sysSize} (h + \var{h}_{j}) \hat{\sigma}_{j}^{(z)},
\end{equation}
where $\hat{\sigma}^{(\alpha)}$ ($\alpha = 1,2,3$) are the Pauli operators, $\var{h}_{j}$ are i.i.d.\ random variables which are uniformly sampled from $[-W, W]$, with $W$ controlling the disorder strength, and $(J_{x}, J_{y}, J_{z}) = (1, 0.5, 1.5)$.
We consider three regimes: (a) nonintegrable ($h = 1$, $W = 0$, periodic boundary), (b) integrable ($h = 0$, $W = 0$, periodic boundary), and (c) many-body localized ($h = 1$, $W = 8$, open boundary).
For calculations under periodic boundary conditions, we restrict ourselves to the zero-momentum and even-parity sector.
The test observables are chosen as $\hat{M}_{z} \coloneqq \sum_{j=1}^{\sysSize} \hat{\sigma}_{j}^{(3)}$, and $C_{xx}$, $C_{yy}$, and $C_{zz}$, each defined as $\sum_{j=1}^{\sysSize} \hat{\sigma}_{j}^{(\alpha)} \hat{\sigma}_{j+1}^{(\alpha)}$ for $\alpha = 1,2,3$.
As shown in Fig.~\ref{fig_XYZ_ETHmeasure_vsN}, $\ETHmeasure{\hat{H}}{\Bqty*{\hat{A}}}$ decays exponentially in (a) and saturates in (b) and (c), consistent with known ETH behavior.
These results confirm both the validity and the sensitivity of our ETH measure for the case of a single observable.
Furthermore, the results in Fig.~\ref{fig_XYZ_ETHmeasure_vsN} provide a valid lower bound on the ETH measure for any operator set $\opSet$ that includes the tested operators, since $\ETHmeasure{\hat{H}}{\opSet} = \sup_{\hat{A} \in \opSet} \ETHmeasure{\hat{H}}{\{\hat{A}\}}$.
In particular, Fig.~\ref{fig_XYZ_ETHmeasure_vsN} offers direct evidence for the breakdown of the ETH with respect to such operator sets in both integrable and many-body localized systems.

\clearpage
\sectionS{Proof of Conjecture~\ref{conj_Conjecture} \protect\\for an idealized pseudo random-matrix model}
\label{SectionSM3}
\subsection*{Model}
As an idealized model of complex many-body eigenstates,
we consider the shell-wise random model repeatedly used in the literature~\citeSM{von2010proofSM, reimann2015generalizationSM}.
It is used even recently: for example, it is called the model of local rotational invariance in Ref.~\citeSM{Pappalardi_PhysRevLett2022zSM}.

Let $\hat{H}$ be the Hamiltonian of a (chaotic) nonintegrable system.
We divide its energy spectrum into disjoint energy shells, whose width is sufficiently small so that the energy dependence within each shell is negligible, while sufficiently large so that each shell contains exponentially many energy levels.
Let $\mathcal{H}_{j}$ denote the $j$th energy shell centered at energy $E_j$ with width $2 \delta_j$,
\begin{equation}
    \mathcal{H}_{j} \coloneqq \vecspan\Bqty{ \ket*{E} \colon (\hat{H} - E) \ket*{E} = 0,\ \abs*{ E - E_{j} } \leq \delta_j }.
\end{equation}
The Hilbert space is decomposed as $\mathcal{H} = \bigoplus_{j} \mathcal{H}_{j}$.
We write $D_j \coloneqq \dim\mathcal{H}_j$ and denote by $\hat{\Pi}_j$ the orthogonal projection onto $\mathcal{H}_j$.
In each shell, choose a reference orthonormal basis $\Bqty*{\ket*{j,\mu}}_{\mu=1}^{D_j}$.
Then, energy eigenstates in the $j$th shell are represented as
\begin{equation}
    \ket*{ E_{(j,\mu)} } = \hat{U}^{(j)} \ket*{j,\mu},
\end{equation}
where $\hat{U}^{(j)}$ is a unitary operator acting on $\mathcal{H}_{j}$.
The Hamiltonian is then written as
\begin{equation}
    \hat{H} = \sum_{j} \sum_{\mu=1}^{D_{j}} E_{(j,\mu)} \hat{U}^{(j)} \dyad*{j,\mu} \hat{U}^{(j) \dag}
    \eqqcolon \hat{U} \hat{H}_{0} \hat{U}^{\dag},
\end{equation}
where $\hat{U} = \oplus_{j} \hat{U}^{(j)}$ and $\hat{H}_{0} = \sum_{j} \sum_{\mu=1}^{D_{j}} E_{(j,\mu)} \dyad*{j,\mu}$.
Once the reference basis is chosen, $\hat{U}^{(j)}$ is determined by the Hamiltonian.
Following Refs.~\citeSM{von2010proofSM, reimann2015generalizationSM, Pappalardi_PhysRevLett2022zSM}, we model the statistics of complex many-body eigenstates using the local rotational invariance ansatz.
That is, we treat $\hat{U}^{(j)}$ as a (pseudo-)random unitary operator acting within the energy shell $\mathcal{H}_j$, with a distribution invariant under unitary rotations of the reference basis inside $\mathcal{H}_j$.
Mathematically, this local rotational invariance defines the Haar distribution for each $\hat{U}^{(j)}$.

The microcanonical density operator on the $j$th shell is given by $\hat{\Pi}_{j} / D_j$.
Therefore, the ETH measure is given by
\begin{equation}
    \ETHmeasure{\hat{H}}{\opSet}(E)
    = \max_{ (j,\mu), (k,\nu) \in \mathcal{I}_{\Delta\!{E}}(E)  }
    \seminorm{1}{ \EigStateOp{(j,\mu)(k,\nu)} - \delta_{jk} \delta_{\mu\nu} \frac{\hat{\Pi}_{j}}{D_{j}} },
\end{equation}
where $\mathcal{I}_{\Delta\!{E}}(E) \coloneqq \Bqty*{ (j,\mu) \colon \abs*{ E_{(j,\mu)} - E } \leq \Delta{E} }$ denotes the index set for which we test the ETH.
From Proposition~\ref{Proposition1} in the main text, the ETH measure can be bounded from above by 
\begin{equation}
    \ETHmeasure{\hat{H}}{\opSet}(E) 
    \leq \sqrt{D}\, \ETHmeasure[2]{\hat{H}}{\opSet}(E)
    \leq \max_{ (j,\mu), (k,\nu) \in \mathcal{I}_{\Delta\!{E}}(E)  } \sqrt{D}\, \norm*{ \vec{\Delta}^{(j,\mu)(k,\nu)}_{U} }_{2},
    \label{eq_IneqETHmeasure}
\end{equation}
where $D \coloneqq \dim\mathcal{H}$ and
\begin{equation}
    (\vec{\Delta}^{(j,\mu)(k,\nu)}_{U})_{\ell} \coloneqq \tr\!\bqty{ \hat{\Lambda}_{\ell}^{\dag} \hat{U} \qty(\dyad*{j,\mu}{k,\nu} - \delta_{jk} \delta_{\mu\nu} \frac{\hat{\Pi}_{j}}{D_{j}}) \hat{U}^{\dag} }.
\end{equation}

The proof proceeds in the following four steps.
First, we bound the average of $f^{(j,\mu)(k,\nu)}_{2}(\hat{U})$ from above in Lemma~\ref{lemSM:AverageBound}.
Second, we show the Lipschitz continuity of $\ETHmeasure[2]{\hat{H}}{\opSet}(E)$ in Eq.~\eqref{eq_LipschitzContinuityOfTrace} so that we can apply the concentration inequality~\eqref{eq_ConcentrationSU_Upper} for the Haar measure on $\prod_j \mathbb{SU}(D_j)$.
Third, we apply the union bounds to obtain the high-probability upper bound on $\ETHmeasure[2]{\hat{H}}{\opSet}(E)$, which is Eq.~\eqref{eq_RandomMatrixEstimate} in the main text.
Finally, we calculate the dimension of the $m$-body operator spaces to show that there exists a function $\Phi(\alpha)$ such that $\dim\opSet_{\particleNumber}^{[0,\alpha\particleNumber]} = e^{ \particleNumber \Phi(\alpha) + o(\particleNumber) }$ and $\Phi(\alpha) \to 0$ as $\alpha \to 0$.

\subsection*{Step~1: Bounding the average from above.}
The following lemma gives an upper bound on the average of $\sqrt{D}\, \ETHmeasure[2]{\hat{H}}{\opSet}(E)$.
\begin{framedLemma}{}{AverageBound}
    Let $\{ \hat{\Lambda}_{\ell} \}_{\ell = 1}^{M}$ be an orthonormal basis of $\opSet + \mathbb{R} \hat{I}$ with $M \coloneqq \dim(\opSet + \mathbb{R} \hat{I})$.
    Suppose that $\sum_{\ell = 1}^{M} \hat{\Lambda}_{\ell}^{\dag} \hat{\Lambda}_{\ell} = (M / D) \hat{I}$.
    (This assumption is fulfilled by the $m$-body operator spaces introduced in the main text.)
    Then, the average of $\sqrt{D}\, \ETHmeasure[2]{\hat{H}}{\opSet}(E)$ is bounded above as
    \begin{equation*}
        \mathbb{E}_{U}\!\bqty{ \sqrt{D}\, \ETHmeasure[2]{\hat{H}}{\opSet}(E) }
        \leq \sqrt{ \frac{ M }{ D_{\mathrm{min}}(E) } } \qty\Big(1 + \order{ D_{\mathrm{min}}(E)^{-1} }),
    \end{equation*}
    where $D_{\mathrm{min}}(E) \coloneqq \displaystyle\min_{j\colon \mathcal{H}_j \cap \mathcal{H}_{E,\Delta\!{E}} \neq \emptyset} D_j$ is the minimum shell dimension overlapping with the ETH test region $\mathcal{H}_{E,\Delta\!{E}}$.
\end{framedLemma}
\begin{proof}
    From Jensen’s inequality, we have
    \begin{equation}
        \mathbb{E}_{U} \norm{ \vec{\Delta}^{(j,\mu)(k,\nu)}_{U} }_{2} 
        \leq \sqrt{ \mathbb{E}_{U}\!\bqty{ \norm{ \vec{\Delta}^{(j,\mu)(k,\nu)}_{U} }_{2}^2 } }.
        \label{eq_Jansen}
    \end{equation}
    Below, we estimate the second moment on the right-hand side.

    For $j \neq k$, we have $(\vec{\Delta}^{(j,\mu)(k,\nu)}_{U})_{\ell} = \matrixel*{ \psi^{(k)} }{ \hat{\Lambda}_{\ell}^{\dag} }{ \psi^{(j)} }$, where $\ket*{\psi^{(j)}} \coloneqq U^{(j)} \ket*{j,\mu}$ and $\ket*{\psi^{(k)}} \coloneqq U^{(k)} \ket*{k,\nu}$ are uniformly distributed vectors on $\mathcal{H}_j$ and $\mathcal{H}_k$, respectively.
    Using the average formula $\mathbb{E}_{U}[ \dyad*{\psi^{(j)}} ] = \Pi_j / D_j$, the second moment is calculated as
    \begin{align}
        \mathbb{E}_{U}\!\bqty{ \norm{ \vec{\Delta}^{(j,\mu)(k,\nu)}_{U} }_{2}^{2} }
        &= \sum_{\ell=1}^{M} \mathbb{E}_{U} \abs{ \matrixel*{ \psi^{(k)} }{ \hat{\Lambda}_{\ell}^{\dag} }{ \psi^{(j)} } }^2
        \nonumber \\
        &= \sum_{\ell=1}^{M} \mathbb{E}_{U^{(k)}} \expval{ \hat{\Lambda}_{\ell}^{\dag} \frac{ \hat{\Pi}_j }{ D_j } \hat{\Lambda}_{\ell} }{ \psi^{(k)} }
        = \sum_{\ell=1}^{M} \frac{ \tr(\hat{\Lambda}_{\ell}^{\dag} \hat{\Pi}_j \hat{\Lambda}_{\ell} \hat{\Pi}_k) }{ D_j D_k }
        \nonumber \\
        &= \sum_{\ell=1}^{M} \frac{ \norm*{ \hat{\Pi}_j \hat{\Lambda}_{\ell} \hat{\Pi}_k }_{2}^{2} }{ D_j D_k }.
    \end{align}
    Applying Hölder's inequality, we obtain
    \begin{equation}
        \sum_{\ell=1}^{M} \norm*{ \hat{\Pi}_j \hat{\Lambda}_{\ell} \hat{\Pi}_k }_{2}^{2} 
        \leq \sum_{\ell=1}^{M} \norm*{ \hat{\Lambda}_{\ell} \hat{\Pi}_k }_{2}^{2}
        = \tr\qty( \hat{\Pi}_k \sum_{\ell=1}^{M} \hat{\Lambda}_{\ell}^{\dag} \hat{\Lambda}_{\ell} \hat{\Pi}_k )
        = \frac{M}{D} D_k,
    \end{equation}
    where we use the assumption $\sum_{\ell=1}^{M} \hat{\Lambda}_{\ell}^{\dag} \hat{\Lambda}_{\ell} = (M / D) \hat{I}$ in deriving the first equality.
    Therefore, we obtain
    \begin{equation}
        \mathbb{E}_{U}\!\bqty{ \norm{ \vec{\Delta}^{(j,\mu)(k,\nu)}_{U} }_{2}^{2} } \leq \frac{M}{D} \frac{1}{ D_j } \quad (j\neq k).
        \label{eq_AverageOffDiag}
    \end{equation}

    For $j = k$, we have
    \begin{equation}
        (\vec{\Delta}^{(j,\mu)(k,\nu)}_{U})_{\ell}
        = \matrixel*{j,\nu}{ \hat{U}^{(j) \dag} \qty( \hat{\Lambda}_{\ell}^{\dag} -  \frac{\tr[\hat{\Lambda}_{\ell}\hat{\Pi}_j]^{\ast}}{D_j} \hat{\Pi}_j ) \hat{U}^{(j)} }{j,\mu}
        = \tr\qty( \hat{U}^{(j)} \hat{E}^{(j)}_{\mu\nu} \hat{U}^{(j) \dag} \mathring{\Lambda}_{\ell}^{\dag} ).
    \end{equation}
    Here, we introduce $\hat{E}^{(j)}_{\mu\nu} \coloneqq \dyad*{j,\mu}{j,\nu}$ and $\mathring{\Lambda}_{\ell} \coloneqq \hat{\Pi}_j \hat{\Lambda}_{\ell} \hat{\Pi}_j - (\tr[\hat{\Lambda}_{\ell} \hat{\Pi}_j]) \hat{\Pi}_j / D_j$.
    Here, we write
    \begin{align}
        \abs*{ (\vec{\Delta}^{(j,\mu)(k,\nu)}_{U})_{\ell} }^2 
        &= \tr\qty( \hat{U}^{(j)} \hat{E}^{(j)}_{\mu\nu} \hat{U}^{(j) \dag} \mathring{\Lambda}_{\ell}^{\dag} ) \tr\qty( \mathring{\Lambda}_{\ell} \hat{U}^{(j)} \hat{E}^{(j) \dag}_{\mu\nu} \hat{U}^{(j) \dag}  )
        \nonumber \\
        &= \tr_{\mathcal{H}^{\otimes 2}}\!\bqty{ (\hat{U}^{(j)})^{\otimes 2} (\hat{E}^{(j)}_{\mu\nu} \otimes \hat{E}^{(j) \dag}_{\mu\nu}) (\hat{U}^{(j) \dag})^{\otimes 2} (\mathring{\Lambda}_{\ell}^{\dag} \otimes \mathring{\Lambda}_{\ell}) }.
    \end{align}
    
    To calculate the second moment, we use the second moment formula~\citeSM{Mele2024-jgSM}:
    \begin{equation}
        \mathbb{E}_{U}[ (\hat{U}^{(j) \dag})^{\otimes 2} \hat{O}^{(j)} (\hat{U}^{(j)})^{\otimes 2} ] \\
        = \frac{D_j \tr \hat{O}^{(j)} - \tr[\hat{O}^{(j)} \mathbb{F}_{2}] }{ D_j (D_j^2 - 1) } \operatorname{Id}_2^{(j)} + \frac{D_j \tr[\hat{O}^{(j)} \mathbb{F}_{2}] - \tr \hat{O}^{(j)} }{ D_j (D_j^2 - 1) } \mathbb{F}_{2}^{(j)},
    \end{equation}
    where $\operatorname{Id}_2^{(j)}$ and $\mathbb{F}_{2}^{(j)}$ denote the identity and the swap operators, respectively, acting on $\mathcal{H}_j^{\otimes 2}$.
    Substituting $\hat{O}^{(j)} = \hat{E}^{(j)}_{\mu\nu} \otimes \hat{E}^{(j) \dag}_{\mu\nu}$, we obtain
    \begin{equation}
        \mathbb{E}_{U}\!\bqty{ \norm{ \vec{\Delta}^{(j,\mu)(j,\nu)}_{U} }_{2}^{2} } 
        = \frac{ D_j - \delta_{\mu\nu} }{ D_j (D_j^2 - 1) } \sum_{\ell=1}^{M} \norm*{ \mathring{\Lambda}_{\ell} }_{2}^{2}.
    \end{equation}
    Here, we have
    \begin{equation}
        \sum_{\ell=1}^{M} \norm*{ \mathring{\Lambda}_{\ell} }_{2}^{2}
        = \sum_{\ell=1}^{M} \qty( \norm*{ \hat{\Pi}_j \hat{\Lambda}_{\ell} \hat{\Pi}_j }_{2}^{2} - \frac{ \abs{ \tr\qty(\hat{\Lambda}_{\ell}^{\dag} \hat{\Pi}_j) }^2 }{D_j} )
        \leq \frac{M}{D} D_j.
    \end{equation}
    Therefore, we obtain
    \begin{equation}
        \mathbb{E}_{U}\!\bqty{ \norm{ \vec{\Delta}^{(j,\mu)(j,\nu)}_{U} }_{2}^{2} }
        \leq \frac{M}{D} \frac{ D_j - \delta_{\mu\nu} }{ D_j^2 - 1 }
        \leq \frac{M}{D} \frac{1}{D_j} \qty( 1 + \order{D_j^{-1}} )
        \label{eq_AverageDiag}
    \end{equation}
    Combining Eqs.~\eqref{eq_AverageOffDiag} and \eqref{eq_AverageDiag}, we obtain
    \begin{equation}
        \mathbb{E}_{U}\!\bqty{ \norm{ \vec{\Delta}^{(j,\mu)(k,\nu)}_{U} }_{2}^{2} }
        \leq \frac{M}{D} \frac{1}{D_j} \qty( 1 + \order{D_j^{-1}} ).
    \end{equation}
    Using Jensen's inequality~\eqref{eq_Jansen} and taking the maximum over the pair of indices $(j,\mu)$ and $(k,\nu)$, we obtain the desired result.
\end{proof}

Lemma~\ref{lemSM:AverageBound} and Eq.~\eqref{eq_IneqETHmeasure} show that the average of the ETH measure vanishes in the thermodynamic limit as long as the ratio $M / D_{\mathrm{min}}(E)$ does so in the same limit.
In the next step, we show that these small deviations occur for almost all realizations of $\hat{U} = \bigoplus_{j} \hat{U}^{(j)}$ for each 
\begin{equation}
    f^{(j,\mu)(k,\nu)}_{1}(\hat{U}) \coloneqq \seminorm{1}{ \hat{U} \qty( \dyad*{j,\mu}{k,\nu} - \delta_{jk} \delta_{\mu\nu} \frac{\hat{\Pi}_j}{D_j} ) \hat{U}^{\dag} }.
\end{equation}

\subsection*{Step~2: Concentration of each $f^{(j,\mu)(k,\nu)}_{1}(\hat{U})$.}
Here, we derive an inequality for the ETH measure $\ETHmeasure{\hat{H}}{\opSet} = \displaystyle\max_{ (j,\mu), (k,\nu) \in \mathcal{I}_{\Delta\!{E}}(E)  } f^{(j,\mu)(k,\nu)}_{1}(\hat{U})$, instead of that for its average.
To do so, we employ the concentration inequality for the Haar measure~\citeSM{Ledoux2001SM,meckes2019randomSM, Mele2024-jgSM}, which applies to any Lipschitz continuous function of $\hat{U}$.
We here show the Lipschitz continuity of $f^{(j,\mu)(k,\nu)}_{1}(\hat{U})$ before introducing the concentration inequality.

For this purpose, we show the Lipschitz continuity of the inner product $\tr(\hat{A} \hat{U} \hat{X} \hat{U}^{\dagger})$ for any $\hat{A}$ and $\hat{X}$ independent of $\hat{U}$. 
This follows from the triangle and the Hölder inequalities as
\begin{align}
    \abs{ \tr(\hat{A} \hat{U}_{1} \hat{X} \hat{U}_{1}^{\dagger}) - \tr(\hat{A} \hat{U}_{2} \hat{X} \hat{U}_{2}^{\dagger}) }
    &\leq \abs{ \tr[ \hat{A} (\hat{U}_{1} - \hat{U}_{2}) \hat{X} \hat{U}_{1}^{\dagger} ] } + \abs{ \tr[ \hat{A} \hat{U}_{2} \hat{X} (\hat{U}_{1}^{\dagger} - \hat{U}_{2}^{\dagger}) ] }
    \nonumber \\
    &\leq 2 \norm*{ \hat{A} }_{\infty} \norm*{ \hat{X} }_{2} \norm*{ \hat{U}_{1} - \hat{U}_{2} }_{2}.
    \label{eq_LipschitzContinuityOfTrace}
\end{align}
Since $\abs*{ \max_{x} g(x) - \max_{x} h(x) } \leq \max_{x} \abs*{ g(x) - h(x) }$ for any functions $g$ and $h$, we conclude from Eq.~\eqref{eq_LipschitzContinuityOfTrace} that $f^{(j,\mu)(k,\nu)}_{1}(\hat{U})$ is Lipschitz continuous with the Lipschitz constant no larger than $2$.
We note that this upper bound is independent of the indices $(j,\mu)$ and $(k,\nu)$.

We can now apply the concentration inequalities for the Haar measure on $\prod_{j} \mathbb{SU}(D_j)$~\citeSM{Ledoux2001SM,meckes2019randomSM, Mele2024-jgSM} to $f^{(j,\mu)(k,\nu)}_{1}(\hat{U})$.
It states that for any $\delta  > 0$ that can depend on $D_j$ and any Lipschitz continuous function $f$ on $\prod_{j} \mathbb{SU}(D_j)$ with Lipschitz constant $\eta_{f}$, the following inequalities hold:
\begin{equation}
    \Prob\bqty{ f(\hat{U}) \geq \mathbb{E}[f] + \delta } \leq \exp\!\bqty{ -\frac{ \delta^2 D_{\mathrm{min}}(E) }{ 4\eta_{f}^2 } }.
    \label{eq_ConcentrationSU_Upper}
\end{equation}
Here, $D_{\mathrm{min}}(E) \coloneqq \displaystyle\min_{j\colon \mathcal{H}_j \cap \mathcal{H}_{E,\Delta\!{E}} \neq \emptyset} D_j$ denotes the minimum shell dimension overlapping with the ETH test region $\mathcal{H}_{E,\Delta\!{E}}$.
This step provides uniform control over all observables in $\mathcal{A}$, going beyond a straightforward application of Levy’s lemma.

\subsection*{Step~3: Controlling the maximum over $(j,\mu)$ and $(k,\nu)$.}
The upper bound in Lemma~\ref{lemSM:AverageBound} and the Lipschitz constant of $f^{(j,\mu)(k,\nu)}_{1}$ are both independent of the indices $(j,\mu)$ and $(k,\nu)$.
Therefore, taking the union bound and then applying Eq.~\eqref{eq_ConcentrationSU_Upper} to each $f^{(j,\mu)(k,\nu)}_{1}$, we obtain
\begin{align}
    \Prob\bqty{ \ETHmeasure{\hat{H}}{\opSet} \geq \sqrt{ \frac{ \dim(\opSet + \mathbb{R} \hat{I}) }{D_{\mathrm{min}}(E)} } \qty( 1 + \order{ D_{\mathrm{min}}^{-1} } ) + \delta }
    &\leq \sum_{(j,\mu), (k,\nu) \in \mathcal{I}_{\Delta\!{E}}(E)} \Prob\bqty{ f^{(j,\mu)(k,\nu)}_{1} \geq \mathbb{E}_{U}[f^{(j,\mu)(k,\nu)}_{1}] + \delta }
    \nonumber \\
    &\leq \abs{\mathcal{I}_{\Delta\!{E}}(E)}^2 \exp\!\bqty{ -\frac{ \delta^2 D_{\mathrm{min}}(E) }{ 4\eta_{f}^2 } }.
    \label{eq_UnionBounds}
\end{align}

Since $\delta > 0$ is arbitrary, we can choose $\delta = \order*{D_{\mathrm{min}}(E)^{\epsilon} / \sqrt{D_{\mathrm{min}}(E)}}$ for any positive $D_{\mathrm{min}}(E)$-independent constant $\epsilon > 0$.
With this choice of $\delta$, the inequality~\eqref{eq_UnionBounds} implies
\begin{equation}
    \ETHmeasure[1]{\hat{H}}{\opSet} 
    \leq \sqrt{\frac{ \dim(\opSet + \mathbb{R} \hat{I}) }{ D_{\mathrm{min}}(E) }} + \order{ \frac{D_{\mathrm{min}}(E)^{\epsilon}}{\sqrt{D_{\mathrm{min}}(E)}} },
    \label{eq_RMmodelETHmeasureBound}
\end{equation}
for sets of $\hat{U}$ whose probability measure is no smaller than $1 - \abs{\mathcal{I}_{\Delta\!{E}}(E)}^2 \exp\!\bqty{ - \order{D_{\mathrm{min}}(E)^{2\epsilon}} }$.
Here, we have
\begin{equation}
    \abs{\mathcal{I}_{\Delta\!{E}}(E)}^2 e^{ - \order{D_{\mathrm{min}}(E)^{2\epsilon}} }
    \leq e^{ - \order{D_{\mathrm{min}}(E)^{2\epsilon}} +2\log D } = e^{ -\order{ D_{\mathrm{min}}(E)^{2\epsilon} } },
\end{equation}
where we use $\abs{\mathcal{I}_{\Delta\!{E}}(E)} \leq D$ and consider finite-temperature region where $D_{\mathrm{min}}(E)$ is exponentially large in the system size.

\subsection*{Step~4: Dimension estimates of the $m$-body operator spaces.}
It remains to estimate $\dim\opSet$ for the $m$-body operator spaces introduced in the main text.
We show that there exists a system-size-independent function $\Phi(\alpha)$ such that $\Phi(\alpha) \xrightarrow{\alpha \to 0} 0$ and
\begin{equation}
    \dim\opSet_{\particleNumber}^{[0,\alpha\particleNumber]}
    \leq
    e^{\particleNumber\Phi(\alpha)+o(\particleNumber)}
    \quad \text{for any fixed $\alpha>0$.}
\end{equation}
Together with the Boltzmann formula $D_{\mathrm{min}}(E)=e^{\particleNumber s_{\mathrm{min}}+o(\particleNumber)}$, where $s_{\mathrm{min}}>0$ is the minimum entropy per particle in the (finite-temperature) ETH test region $\mathcal{H}_{E,\Delta\!{E}}$, this implies the existence of a nonzero constant $\lowerBound^{\text{RM}}$ such that, for any $\alpha<\lowerBound^{\text{RM}}$, Eq.~\eqref{eq_RMmodelETHmeasureBound} gives
\begin{equation}
    \ETHmeasure[1]{\hat{H}}{\opSet_{\particleNumber}^{[0,\alpha\particleNumber]}}
    \leq
    e^{\frac{\particleNumber}{2}\qty[\Phi(\alpha)-s_{\mathrm{min}}]+o(\particleNumber)} + e^{-(1/2-\epsilon)\particleNumber s_{\mathrm{min}}+o(\particleNumber)}
    \xrightarrow{\particleNumber\to\infty}0
\end{equation}
for a set of $\hat U$ whose probability measure is no smaller than $1-\exp[-\order{D_{\mathrm{min}}(E)^{2\epsilon}}]$.
(Here, we choose $0 < \epsilon < 1/2$.)
This proves Conjecture~\ref{conj_Conjecture} for the (pseudo)random-matrix model considered in this section.

\textbf{Spin systems.---}
For $\particleNumber$-site spin-$S$ systems, the operators in $\opSet_{\particleNumber}^{[0,m]}$ are generated by choosing $m'$ sites and a nonidentity local basis operator on each chosen site. 
Hence,
\begin{equation}
    \dim\opSet_{\particleNumber}^{[0,m]} = \sum_{m'=0}^{m} \binom{\particleNumber}{m'} \qty(\dimLoc^{2}-1)^{m'},
\end{equation}
where $\dimLoc = 2S+1$.
We use the elementary bound $\binom{\particleNumber}{m'} \leq e^{ \particleNumber \mathrm{H}(m'/\particleNumber) }$ with the binary entropy $\mathrm{H}(x)\coloneqq -x\log x-(1-x)\log(1-x)$.
This gives 
\begin{equation}
    \dim\opSet_{\particleNumber}^{[0,\alpha \particleNumber]} 
    \leq (\alpha \particleNumber+1) \max_{0 \leq m' \leq \alpha\particleNumber} \binom{\particleNumber}{m'} \qty(\dimLoc^{2}-1)^{m'}
    = e^{ \particleNumber \Phi_{\mathrm{spin}}(\alpha) + \order{\log \particleNumber} },
\end{equation}
where $\Phi_{\mathrm{spin}}(\alpha) \coloneqq \max_{0 \leq x \leq \alpha}\{ \mathrm{H}(x) + x \log(\dimLoc^2 - 1) \}$.
Here, $\Phi_{\mathrm{spin}}(\alpha) \to 0$ as $\alpha \to 0$ is obvious from definition.

\textbf{Bose systems.---}
For $\particleNumber$-particle Bose systems on $\sysSize$ lattices, the creation indices and annihilation indices are multisets. 
Therefore, the operator space dimension is given by
\begin{equation}
    \dim\opSet_{\particleNumber, \sysSize}^{[0,m]} = \binom{\sysSize+m-1}{m}^{2}.
\end{equation}
The Stirling formula gives
\begin{equation}
    \dim\opSet_{\particleNumber, \sysSize}^{[0,\alpha \particleNumber]}
    = e^{ 2 (\sysSize+\alpha \particleNumber-1) \mathrm{H}\qty( \frac{\alpha \particleNumber}{\sysSize+\alpha \particleNumber-1} ) + \order{\log \particleNumber} }
    = e^{ 2 \sysSize (1 + n \alpha) \mathrm{H}\qty( \frac{n \alpha}{1 + n \alpha} ) + o(\particleNumber) },
\end{equation}
where $n \coloneqq \particleNumber / \sysSize$ is the filling factor.
Here, we have
\begin{align}
    \Phi_{\mathrm{Bose}}(\alpha) 
    &\coloneqq 2 \frac{1 + n \alpha}{n} \mathrm{H}\qty( \frac{n \alpha}{1 + n \alpha} )
    \nonumber \\
    &= -2 \alpha \log\frac{n \alpha}{1 + n \alpha} - \frac{2}{n} \log\frac{1}{1 + n \alpha}
    \nonumber \\
    &\xrightarrow{\alpha \to 0} 0.
\end{align}

\textbf{Fermi systems.---}
To calculate the dimension of $\opSet_{N,V}^{[0,m]}$, we consider the particle-hole transformation $\hat{C}$ defined by $\hat{C} \hat{f}_{x} \hat{C}^{\dagger} = \hat{f}_{x}^{\dagger}$ and $\hat{C} \ket*{0} = \hat{f}_{1}^{\dagger} \cdots \hat{f}_{\sysSize}^{\dagger} \ket*{0}$.
Then, we have
\begin{equation}
    \hat{C} \hat{\Pi}_{\particleNumber, \sysSize}
    ( \hat{f}_{x_{1}}^{\dagger} \cdots \hat{f}_{x_{m}}^{\dagger} \hat{f}_{y_{1}} \cdots \hat{f}_{y_{m}} )
    \hat{\Pi}_{\particleNumber, \sysSize} \hat{C}^{\dagger}
    = \hat{\Pi}_{\sysSize - \particleNumber, \sysSize}
    ( \hat{f}_{x_{1}} \cdots \hat{f}_{x_{m}} \hat{f}_{y_{1}}^{\dagger} \cdots \hat{f}_{y_{m}}^{\dagger} )
    \hat{\Pi}_{\sysSize - \particleNumber, \sysSize}.
    \label{eq_ParticleHoleOperation}
\end{equation}
Since exchanging the order of operators $\hat{f}_{x_{1}} \cdots \hat{f}_{x_{m}}$ and $\hat{f}_{y_{1}}^{\dagger} \cdots \hat{f}_{y_{m}}^{\dagger}$ results in additional terms of at most $(m-1)$-body ones, Eq.~\eqref{eq_ParticleHoleOperation} implies $\dim \opSet_{\particleNumber, \sysSize}^{[0,m]} = \dim(\hat{C} \opSet_{\particleNumber, \sysSize}^{[0,m]} \hat{C}^{\dagger}) \leq \dim \opSet_{\sysSize - \particleNumber, \sysSize}^{[0,m]}$ for $m \leq \min\Bqty*{\particleNumber, \sysSize - \particleNumber}$.
By setting $\particleNumber \mapsto \sysSize - \particleNumber$ in the above argument, we obtain the reversed inequality.
Therefore, we arrive at 
\begin{equation}
    \dim \opSet_{\particleNumber, \sysSize}^{[0,m]} = \dim \opSet_{\sysSize - \particleNumber, \sysSize}^{[0,m]}
    \label{eq_ParticleHoleDim}
\end{equation}
for $m \leq \min\!\Bqty*{\particleNumber, \sysSize-\particleNumber}$.
This relation combined with the inclusion relation $\opSet_{N,V}^{[0,m-1]} \subset \opSet_{N,V}^{[0,m]} \subset \mathcal{L}(\mathcal{H})$ gives $\dim \opSet_{\particleNumber, \sysSize}^{[0,m]} = D^{2}$ for $m \geq \min\Bqty*{\particleNumber, \sysSize - \particleNumber}$.
Because of Eq.~\eqref{eq_ParticleHoleDim}, the remaining task is to calculate $\dim \opSet_{\particleNumber, \sysSize}^{[0,m]}$ for $\particleNumber / \sysSize < 1/2$.
In this case, there are $\binom{\sysSize}{m}$ choices for both $\Bqty*{ x_{1}, \cdots, x_{m} }$ and $\Bqty*{ y_{1}, \cdots, y_{m} }$ to have nonzero basis operators.
We can show that different choices for these indices give linearly independent basis operators~(see Sec.~\ref{SectionSM4} for the proof).
Therefore, the dimension of the $(\alpha \particleNumber)$-body operator space is calculated to be 
\begin{equation}
    \dim \opSet_{\particleNumber, \sysSize}^{[0,\alpha \particleNumber]} 
    = \binom{ \sysSize }{ \min\!\Bqty*{\alpha \particleNumber, \sysSize - \particleNumber} }^2
    = e^{ 2 \sysSize \mathrm{H}\qty( \frac{ \min\!\Bqty*{\alpha \particleNumber, \sysSize - \particleNumber} }{ \sysSize } ) + \order{\log \particleNumber} }.
\end{equation}
Here, we have
\begin{equation}
    \Phi_{\mathrm{Fermi}}(\alpha) 
    \coloneqq \frac{2}{n}\, \mathrm{H}\qty( \min\!\Bqty*{n \alpha, 1 - n} )
    \xrightarrow{\alpha \to 0} 0.
\end{equation}

\clearpage
\sectionS{Linear independence of fermionic $m$-body operators}
\label{SectionSM4}
In this section, we show the linear independence of the basis operators 
\begin{equation}
    \hat{f}_{\vb{x}}^{\dagger} \hat{f}_{\vb{y}} 
    \coloneqq \hat{\Pi}_{\particleNumber,\sysSize} \qty( \hat{f}_{x_{m}}^{\dagger} \cdots \hat{f}_{x_{1}}^{\dagger} \hat{f}_{y_{1}} \cdots \hat{f}_{y_{m}} ) \hat{\Pi}_{\particleNumber,\sysSize}\qc
    \qty(
    \begin{aligned}
        1 \leq x_{1} < \cdots < x_{m} \leq \sysSize \\[0px]
        1 \leq y_{1} < \cdots < y_{m} \leq \sysSize
    \end{aligned}
    )
\end{equation}
of the fermionic $m$-body operator space, where $\vb{x} \coloneqq \Bqty*{ x_{1}, \cdots, x_{m} }$ and $\vb{y} \coloneqq \Bqty*{ y_{1}, \cdots, y_{m} }$.
We decompose $\vb{x}$ and $\vb{y}$ as $\vb{x} = X\sqcup Z$ and $\vb{y} = Y\sqcup Z$, where \enquote{$\sqcup$} denotes the disjoint union, $Z \coloneqq \vb{x} \cap \vb{y}$, $X \coloneqq \vb{x}\setminus Z$, and $Y \coloneqq \vb{y} \setminus Z$.
Accordingly, we rearrange the product $\hat{f}_{y_{1}} \cdots \hat{f}_{y_{m}}$ by introducing
\begin{equation}
    \hat{f}_{Y\sqcup Z} \coloneqq \qty( \prod_{x\in Y} \hat{f}_{x} ) \qty( \prod_{x\in Z} \hat{f}_{x} ),
\end{equation}
where the product $\prod$ of the annihilation operators $\hat{f}_{x}$ is arranged in ascending order in $x$ within each pair of parentheses.

To show the linear independence of the operators 
$
    \Bqty{
        \hat{f}_{\vb{x}}^{\dagger} \hat{f}_{\vb{y}}
        \mid 
        \substack{
            1 \leq x_{1} < \cdots < x_{m} \leq \sysSize,\ \\[1ex] 
            1 \leq y_{1} < \cdots < y_{m} \leq \sysSize 
        }
    },
$
we consider the equation
\begin{align}
    0 = 
    \sum_{ \substack{ X,Y \\ X\cap Y = \emptyset \\ \abs*{X} = \abs*{Y} \leq \min\Bqty{ m, \sysSize/2 } }} 
    \sum_{\substack{ Z \\ \abs*{Z} = m - \abs*{X} \\ Z\cap X = \emptyset, Z\cap Y = \emptyset }}
    C_{X,Y}(Z) ( \hat{f}_{X\sqcup Z} )^{\dagger} \hat{f}_{Y\sqcup Z}
    \quad \qty( \eqqcolon \hat{A} ).
    \label{eqS_LinearIndependence}
\end{align}

Given $X$ and $Y$, we label the basis vectors of $\mathcal{H}_{\particleNumber, \sysSize}$ as
\begin{equation}
    \ket*{ \vb{n}_{X}, \vb{n}_{Y}, \vb{n}_{ \vb{V} \setminus (X\sqcup Y) } } 
    \coloneqq \qty( \prod_{ x\in X } (\hat{f}_{x} )^{n_{x}} )^{\dagger} \qty( \prod_{ y\in Y } (\hat{f}_{y} )^{n_{y}} )^{\dagger} \qty( \prod_{ x\in \vb{V} \setminus (X\sqcup Y) } (\hat{f}_{x} )^{n_{x}} )^{\dagger} \ket*{0},
\end{equation}
where $\vb{n}_{W} \coloneqq \Bqty*{ n_{x} \in \Bqty*{0, 1} \mid x\in W }$ with $W$ being an arbitrary subset of $\vb{V} \coloneqq \Bqty*{ 1,\cdots, V }$.
Then, taking the matrix element of the right-hand side of Eq.~\eqref{eqS_LinearIndependence}, we obtain
\begin{align}
    &\quad \matrixel*{ \vb{n}_{X} = \vb{1}, \vb{n}_{Y} = \vb{0}, \vb{n}_{ \vb{V} \setminus (X\sqcup Y) } }{ \hat{A} }{ \vb{n}_{X} = \vb{0}, \vb{n}_{Y} = \vb{1}, \vb{n}_{ \vb{V} \setminus (X\sqcup Y) } }
    \nonumber \\
    &= \sum_{Z} C_{X,Y}(Z) \matrixel{ \vb{n}_{X} = \vb{1}, \vb{n}_{Y} = \vb{0}, \vb{n}_{ \vb{V} \setminus (X\sqcup Y) } }{ \hat{f}_{X\sqcup Z}^{\dagger} \hat{f}_{Y\sqcup Z} }{ \vb{n}_{X} = \vb{0}, \vb{n}_{Y} = \vb{1}, \vb{n}_{ \vb{V} \setminus (X\sqcup Y) } }
    \nonumber \\
    &= \sum_{Z} C_{X,Y}(Z) \matrixel*{ \vb{n}_{ \vb{V} \setminus (X\sqcup Y) } }{ \hat{f}_{Z}^{\dagger} \hat{f}_{Z} }{ \vb{n}_{ \vb{V} \setminus (X\sqcup Y) } }
    \nonumber \\
    &= \sum_{Z} C_{X,Y}(Z) \chi\qty( \forall x\in Z,\ n_{x} = 1 ),
\end{align}
where $\chi(\phi)$ is the indicator function that takes on unity if the proposition $\phi$ is true and zero otherwise.
Therefore, the equation~\eqref{eqS_LinearIndependence} implies
\begin{equation}
    \sum_{\substack{ Z \\ \abs*{Z} = m - \abs*{X} \\ Z\cap X = \emptyset,\, Z\cap Y = \emptyset }} C_{X,Y}(Z) \chi\qty( \forall x\in Z,\ n_{x} = 1 ) = 0,
    \label{eqS_LinearIndependence2}
\end{equation}
which holds for all disjoint subsets $X, Y \subset \vb{V}$ and all configurations $\vb{n}_{\vb{V} \setminus (X \sqcup Y)}$ such that:
\begin{equation}
    X\cap Y = \emptyset\qc
    \abs*{X} = \abs*{Y} \leq m\qc
    \sum_{ x \in \vb{V}\setminus (X\sqcup Y) } n_{x} = \particleNumber - \abs*{X}.
\end{equation}

Here, the range of the summation over $Z$ in Eq.~\eqref{eqS_LinearIndependence2} depends only on $X$ and $Y$ and is independent of $\vb{n}_{ \vb{V} \setminus (X\sqcup Y) }$, and the number of the summation over $Z$ is given by
\begin{equation}
    \binom{ \sysSize - \abs*{X} - \abs*{Y} }{ \abs*{Z} } = \binom{ \sysSize - 2\abs*{X} }{ m - \abs*{X} }.
    \label{eqS_NumberOfCoeffs}
\end{equation}
On the other hand, the number of choices for $\vb{n}_{ \vb{V} \setminus (X\sqcup Y) }$ in Eq.~\eqref{eqS_LinearIndependence2} is 
\begin{equation}
    \binom{ \sysSize - 2\abs*{X} }{ \particleNumber - \abs*{X} }
    \label{eqS_NumberOfEqs}
\end{equation}
for given $X$ and $Y$.
When $\particleNumber \leq \sysSize/2$, we have
\begin{equation}
    m - \abs*{X} \leq \particleNumber - \abs*{X} \leq \frac{\sysSize}{2} - \abs*{X} = \frac{ \sysSize - 2\abs*{X} }{2},
\end{equation}
and therefore
\begin{equation}
    \binom{ \sysSize - 2\abs*{X} }{ m - \abs*{X} } \leq \binom{ \sysSize - 2\abs*{X} }{ \particleNumber - \abs*{X} }
\end{equation}
for any $m\leq \particleNumber$.
This means that the number of equations~\eqref{eqS_LinearIndependence2} for given $X$ and $Y$ is equal to or larger than the number of the coefficients $\Bqty*{ C_{X, Y}(Z) }_{Z}$.
Therefore, Eq.~\eqref{eqS_LinearIndependence} implies $C_{X,Y}(Z) = 0$ for all $X,Y,Z$ when $\particleNumber \leq \sysSize/2$.
Thus, the linear independence of the operators 
$
    \Bqty{
        \hat{f}_{\vb{x}}^{\dagger} \hat{f}_{\vb{y}}
        \mid 
            1 \leq x_{1} < \cdots < x_{m} \leq \sysSize,\ 
            1 \leq y_{1} < \cdots < y_{m} \leq \sysSize 
    }
$
is proved.

On the other hand, when $N > \sysSize/2$, we have
\begin{equation}
    \sysSize - 2\abs*{X} - (\particleNumber - \abs*{X}) = \sysSize - \particleNumber - \abs*{X} \leq \frac{ \sysSize - 2\abs*{X} }{ 2 }.
\end{equation}
Therefore, for $m > \sysSize-\particleNumber$, the number of equations ~\eqref{eqS_LinearIndependence2} for given $X$ and $Y$ is less than the number of the coefficients $\Bqty*{ C_{X,Y}(Z) }_{Z}$, and the operators in
$
    \Bqty{
        \hat{f}_{\vb{x}}^{\dagger} \hat{f}_{\vb{y}}
        \mid 
            1 \leq x_{1} < \cdots < x_{m} \leq \sysSize, 
            1 \leq y_{1} < \cdots < y_{m} \leq \sysSize 
    }
$
can be linearly dependent.

\begin{figure}[h]
    \centering
    \includegraphics[width=\linewidth]{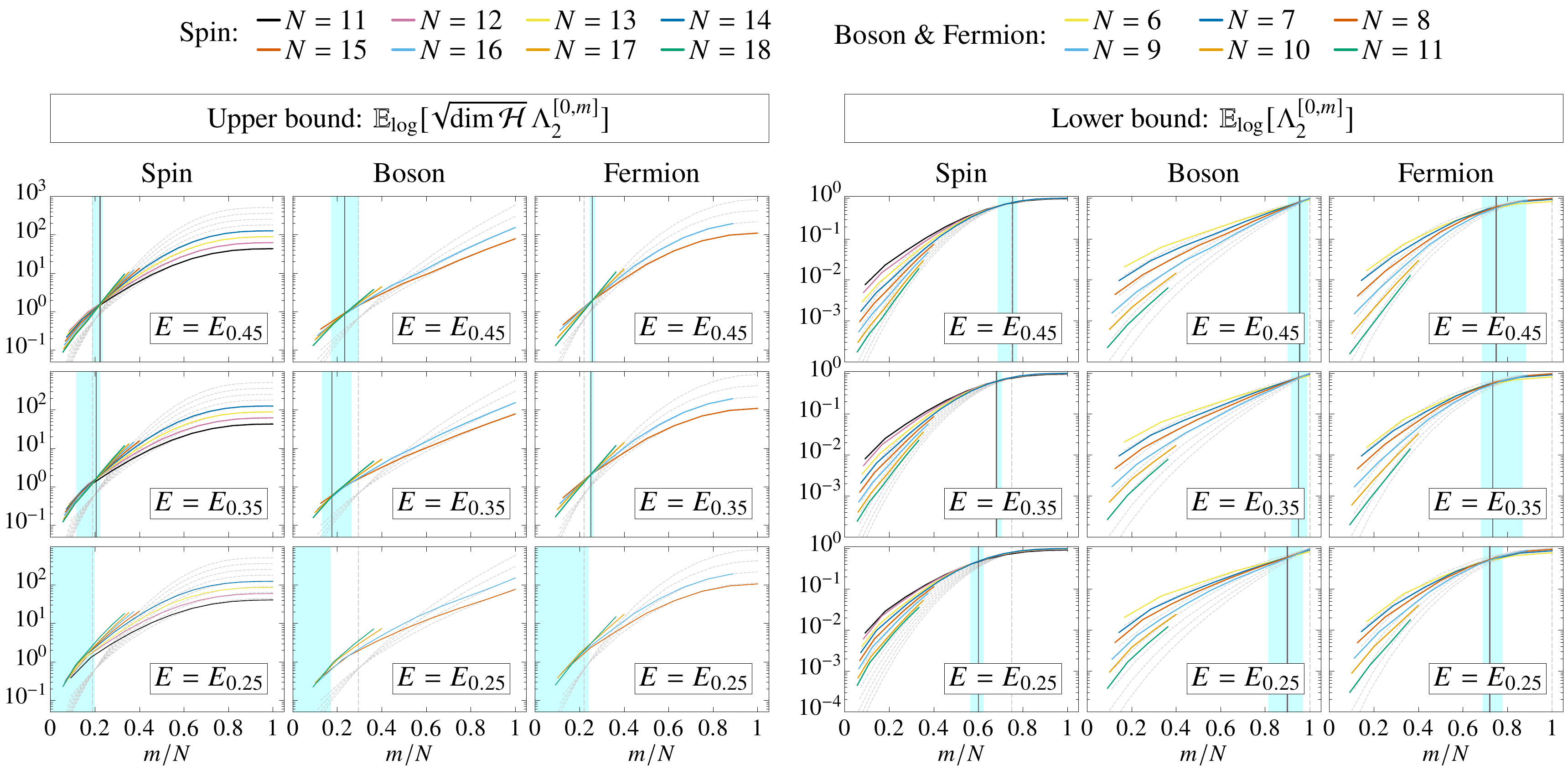}
    \caption{\textbf{Upper and lower bounds on the (diagonal) ETH measure for varying the normalized energy in locally interacting systems.}
    The gray solid vertical line indicates the point $\alpha_{\mathrm{L/U}}{(x)}$ where the maximum difference of the curves for different system sizes is minimized.
    The blue shaded region shows where the $N$-dependence of the bounds ($\mathbb{E}_{\mathrm{log}}[ \sqrt{\dim\mathcal{H}}\, \Lambda_{2}^{[0,\alpha \particleNumber]} ]$ and $\mathbb{E}_{\mathrm{log}}[ \Lambda_{2}^{[0,\alpha \particleNumber]} ]$) qualitatively changes from decreasing~(left) to increasing~(right), indicating the uncertainty of $\lowerBound{}$ and $\upperBound{}$ due to limited system sizes.
    The gray dashed curves and vertical lines show the upper bound in Eq.~\eqref{eq_RandomMatrixEstimate} in the main text derived for an idealized (pseudo)random-matrix model.
    The intersection point $\lowerBound{(x)}$ of $\mathbb{E}_{\mathrm{log}}[ \sqrt{\dim\mathcal{H}}\, \Lambda_{2}^{[0,\alpha \particleNumber]} ]$ is observed for non-small $x$ but not observed for small $x$ ($x \lesssim 0.25$ for spin and Fermi systems and $x \lesssim 0.3$ for Bose systems).
    The intersection point $\upperBound{(x)}$ of $\mathbb{E}_{\mathrm{log}}[ \Lambda_{2}^{[0,\alpha \particleNumber]} ]$ is also observed for all values of $x$ shown in the figure.}
    \label{figS_UpperBound_EnergyDeps}
\end{figure}

\clearpage
\sectionS{Bounds $\sqrt{\dim\mathcal{H}}\, \Lambda_{2}^{[0,m]}$ and $\Lambda_{2}^{[0,m]}$ at various energies}
\label{SectionSM5}
Figure~\ref{fig_Numerics_ShortRangeEnsemble} in the main text presents the numerical results for the upper bound $\sqrt{\dim\mathcal{H}}\, \Lambda_{2}^{[0,m]}$ of the diagonal ETH measure at the center of the energy spectrum, $E_{x} = E_{0.5} = (E_{\mathrm{max}} + E_{\mathrm{min}})/2$, where $\Lambda_{2}^{[0,m]}$ is defined in the main text.
Figure~\ref{fig_BoundsEnergyDependence} in the main text shows the resulting lower and upper bounds on the threshold $\alpha_{\ast}$, $\lowerBound{} \leq \alpha_{\ast} \leq \upperBound{}$ across the energy spectrum.
These bounds are derived from the results for $\sqrt{\dim\mathcal{H}}\, \Lambda_{2}^{[0,m]}$ and $\Lambda_{2}^{[0,m]}$ at the corresponding energies.
These results are presented here in Fig.~\ref{figS_UpperBound_EnergyDeps}.

\clearpage
\sectionS{Numerical results for concrete physical models}
\label{SectionSM6}
In this section, we show the numerical results for the following prototypical models of nonintegrable systems:
\begin{itemize}
    \setlength{\itemsep}{0pt}
    \item Spin: We consider the mixed-field Ising model, whose Hamiltonian is given by
    \begin{equation}
        \hat{H} = J_{x} \sum_{j=1}^{\sysSize} \hat{\sigma}_{j}^{(1)} \hat{\sigma}_{j+1}^{(1)} +B_{z} \sum_{j=1}^{\sysSize} \hat{\sigma}_{j}^{(3)} +B_{x} \sum_{j=1}^{\sysSize} \hat{\sigma}_{j}^{(1)},
        \label{eqS_MixedFieldIsingModel}
    \end{equation}
    where $\hat{\sigma}_{j}^{(p)} \ (p = 1,2,3)$ are Pauli operators acting on site $j$.
    This model is numerically verified to satisfy the ETH in Ref.~\citeSM{kim2014testingSM} for the parameter $(J_{x}, B_{z}, B_{x}) = (1, 0.9045, 0.8090)$.
    \item Boson: We consider the Bose-Hubbard model, whose Hamiltonian is given by
    \begin{align}
        \hat{H} 
        &= -J \sum_{x=1}^{\sysSize} ( \hat{b}_{x}^{\dagger} \hat{b}_{x+1} + \hat{b}_{x+1}^{\dagger} \hat{b}_{x} ) -U \sum_{x=1}^{\sysSize} \frac{ \hat{n}_{x} (\hat{n}_{x} - 1) }{2}.
        \label{eqS_BoseHubbard}
    \end{align}
    This model at unit filling ($\particleNumber/\sysSize = 1$) is numerically verified to satisfy the ETH in Ref.~\citeSM{biroli2010effectSM} for the parameter $(J,U) = (1,1)$, where it is known to be nonintegrable~\citeSM{Kolovsky2004-ooSM}.
    We use these parameters in the numerical calculations.
     \item Fermion: We consider the spinless fermions with next-nearest hopping and interactions:
    \begin{align}
    \hat{H} 
    &= -t_{1} \sum_{x=1}^{\sysSize} ( \hat{f}_{x}^{\dagger} \hat{f}_{x+1} + \hat{f}_{x+1}^{\dagger} \hat{f}_{x} ) -J_{1} \sum_{x=1}^{\sysSize} \hat{n}_{x} \hat{n}_{x+1}
    \nonumber \\
    &\qquad 
    -t_{2} \sum_{x=1}^{\sysSize} ( \hat{f}_{x}^{\dagger} \hat{f}_{x+2} + \hat{f}_{x+2}^{\dagger} \hat{f}_{x} ) -J_{2} \sum_{x=1}^{\sysSize} \hat{n}_{x} \hat{n}_{x+2}.
    \label{eqS_FermiHubbardNNN}
    \end{align}
    This model is numerically verified to exhibit quantum chaotic behavior in Ref.~\citeSM{Santos2010-qsSM} for $(t_{1}, J_{1}, t_{2}, J_{2}) = (1, 1, 0.32, 0.32)$ and $\particleNumber/\sysSize = 1/3$.
    In our numerical calculation, we use these values of $(t_{1}, J_{1}, t_{2}, J_{2})$ but set $\particleNumber/\sysSize = 1/2$.
\end{itemize}
We adopt periodic boundary conditions in all the models listed above.
Since these systems have translation and reflection symmetries, we focus on the zero-momentum and even-parity sector.

\begin{figure}[h]
    \centering
    \includegraphics[width=\linewidth]{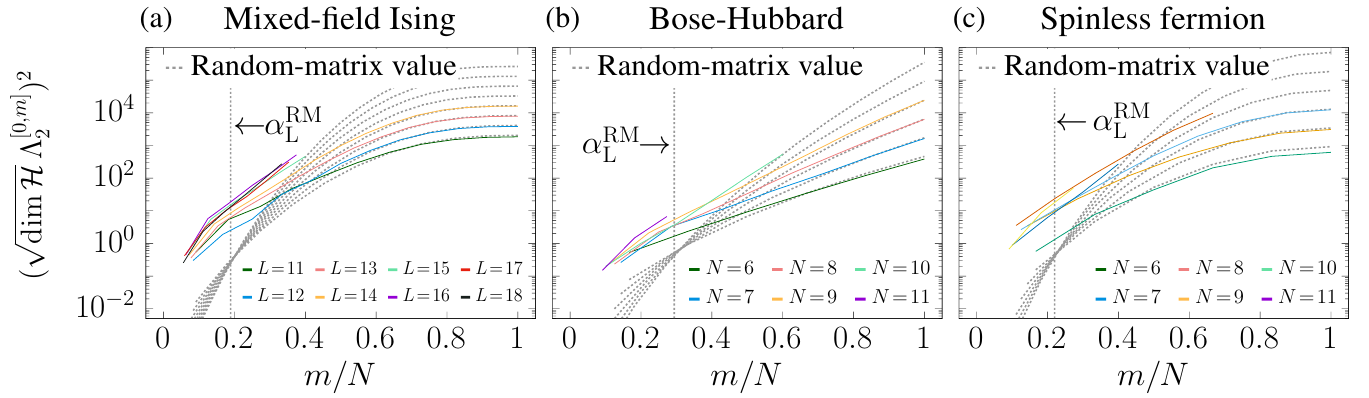}
    \caption{\textbf{Upper bound on the (diagonal) ETH measure as a function of $m/\particleNumber$ for the nonintegrable models in Eqs.~\eqref{eqS_MixedFieldIsingModel}--\eqref{eqS_FermiHubbardNNN}.}
    The upper bounds are calculated at the center of the energy spectrum.
    While some of the curves intersect with each other, they do not intersect at a single point within the system sizes of the numerical calculation.}
    \label{fig_ConcreteModels_UpperBoundVsM}
    \vspace{4ex}
    \centering
    \includegraphics[width=\linewidth]{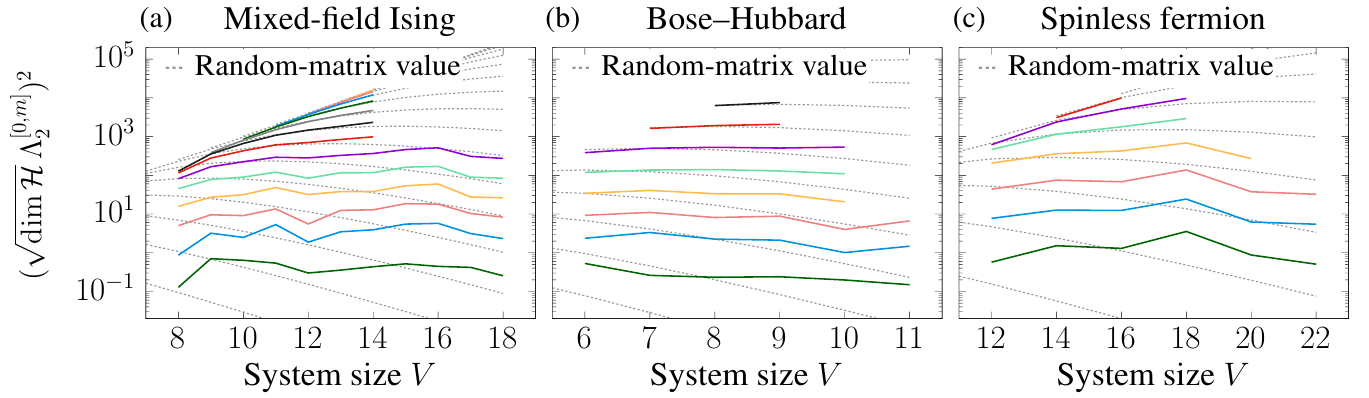}
    \caption{\textbf{Upper bound on the (diagonal) ETH measure as a function of $\sysSize$ for the nonintegrable models in Eqs.~\eqref{eqS_MixedFieldIsingModel}--\eqref{eqS_FermiHubbardNNN}.}
    The same data as in Fig.~\ref{fig_ConcreteModels_UpperBoundVsM} are shown as a function of the system size.
    Different colors indicate different values of $m$ ranging from $1$ to $\particleNumber$.
    The upper bound $\sqrt{\dim\mathcal{H}}\, \Lambda_{2}^{[0,m]}$ with $m\in[1,4]$ decreases as $\sysSize$ increases in the region $\sysSize \geq 16$ for the spin system and $\sysSize \geq 9$ for the spinless fermions.
    However, the number of data points in these regions is too small, and the $\sysSize$-dependence of $\sqrt{\dim\mathcal{H}}\, \Lambda_{2}^{[0,m]}$ is not smooth.
    For the boson system, the upper bound $\sqrt{\dim\mathcal{H}}\, \Lambda_{2}^{[0,m]}$ with $m=1$ smoothly decreases for $\sysSize \geq 6$.
    However, the curves for $m=2,3$ increase as $\sysSize$ increases from $10$ to $11$, which are the two largest system sizes in our calculation.
    These facts indicate that the finite-size effects are nonnegligible in determining the behavior of $\sqrt{\dim\mathcal{H}}\, \Lambda_{2}^{[0,m]}$ for large $\sysSize$ in the nonintegrable models in Eqs.~\eqref{eqS_MixedFieldIsingModel}--\eqref{eqS_FermiHubbardNNN}.}
    \label{fig_ConcreteModels_UpperBoundVsL}
\end{figure}

Figure~\ref{fig_ConcreteModels_UpperBoundVsM} shows the upper bound $\sqrt{\dim\mathcal{H}}\, \Lambda_{2}^{[0,m]}$ on the (diagonal) ETH measure at the center of the energy spectrum as a function of $m/\particleNumber$.
We observe that some of the curves with different system sizes intersect with each other.
However, unlike the case for generic locally interacting systems shown in Fig.~\ref{figS_UpperBound_EnergyDeps}, the point where curves for various system sizes almost intersect is not observed for the nonintegrable models in Eqs.~\eqref{eqS_MixedFieldIsingModel}--\eqref{eqS_FermiHubbardNNN}.
Thus, it remains to be an open problem whether the intersection point converges to a finite value or vanishes in the thermodynamic limit for these models.

It is interesting that the finite-size behaviors of $\sqrt{\dim\mathcal{H}}\, \Lambda_{2}^{[0,m]}$ are different in the nonintegrable models and the generic locally interacting systems presented in the main text, even though these systems share the essential physical structure of the locality and the few-body nature of interactions.
It is an important future problem to investigate the underlying causes of these differences.

Figure~\ref{fig_ConcreteModels_UpperBoundVsL} shows the same data as in Fig.~\ref{fig_ConcreteModels_UpperBoundVsM} as a function of the system size.
For the mixed-field Ising model~\eqref{eqS_MixedFieldIsingModel} and the spinless fermion model~\eqref{eqS_FermiHubbardNNN}, the upper bound $\sqrt{\dim\mathcal{H}}\, \Lambda_{2}^{[0,m]}$ with $m\in[1,4]$ decreases for the three largest system sizes.
However, the number of these data points is too small, and the $\sysSize$-dependence of $\sqrt{\dim\mathcal{H}}\, \Lambda_{2}^{[0,m]}$ is not smooth for the system sizes of the numerical calculation.
For the Bose-Hubbard model~\eqref{eqS_BoseHubbard}, the upper bound $\sqrt{\dim\mathcal{H}}\, \Lambda_{2}^{[0,m]}$ with $m=1$ almost monotonically decreases for $\sysSize \geq 6$.
However, the curves for $m=2,3$ increase as $\sysSize$ increases from $10$ to $11$, which are the two largest system sizes in our calculation.
These facts indicate that the finite-size effects are nonnegligible in determining the behavior of $\sqrt{\dim\mathcal{H}}\, \Lambda_{2}^{[0,m]}$ for large $\sysSize$ in the nonintegrable models in Eqs.~\eqref{eqS_MixedFieldIsingModel}--\eqref{eqS_FermiHubbardNNN}.

The large finite-size effects of $\sqrt{\dim\mathcal{H}}\, \Lambda_{2}^{[0,m]}$ in Figs.~\ref{fig_ConcreteModels_UpperBoundVsM} and \ref{fig_ConcreteModels_UpperBoundVsL} may suggest that the inequality $\seminorm{1}{\cdot} \leq \sqrt{\dim\mathcal{H}}\, \seminorm{2}{\cdot}$ is too loose to test Conjecture~\ref{conj_Conjecture} in the main text for those concrete physical models.
It remains to be an important future task to validate (or invalidate) Conjecture~\ref{conj_Conjecture} for prototypical nonintegrable systems, e.g., by finding better bounds on the ETH measure.

\bibliographySM{supplement}

\end{document}